\documentclass[aps,footinbib,longbibliography, notitlepage,superscriptaddress,eqsecnum]{revtex4-1}
\usepackage{graphicx}
 \usepackage{amssymb}
 \usepackage{amsmath}
 \usepackage{amsfonts}
\usepackage{overpic}
\usepackage{enumerate}
\usepackage{color}
\usepackage[dvipsnames]{xcolor}
\usepackage{xspace}
\usepackage[normalsize]{subfigure}
\usepackage{hyperref}
\usepackage{bm}
\usepackage{empheq}
\usepackage[capitalize]{cleveref}

\crefname{section}{Sec.}{Secs.}

\DeclareFontFamily{OT1}{pzc}{}
\DeclareFontShape{OT1}{pzc}{m}{it}{<-> s * [1.10] pzcmi7t}{}
\DeclareMathAlphabet{\mathpzc}{OT1}{pzc}{m}{it}

\providecommand{\sfrac}[2]{#1/#2}

\providecommand{\ut}[1]{^{\text{#1}}}

\def\onehalf{\frac{1}{2}}

\def\bra{\ensuremath{\langle}}
\def\ket{\ensuremath{\rangle}}

\def\const{\mathrm{const}}

\def\pd{\partial}

\def\im{\mathrm{i}}

\def\qv{\bv{q}}

\def\vv{\bv{v}}
\def\pv{\bv{p}}
\def\rv{\bv{r}}

\def\rvp{\bv{r}_\parallel}

\def\Rv{\bv{R}}

\def\b0{\bv{0}}

\def\Hcal{\mathcal{H}}
\def\Hc2{\Hcal^{(2)}}

\def\Dcal{\mathcal{D}}
\def\Kcal{\mathcal{K}}

\def\Ocal{\mathcal{O}}

\def\Zcal{\mathcal{Z}}

\def\hyp13{{_1 F_3}}

\def\bcs{BCs\xspace}
\def\pbc{\ut{(p)}}
\def\Pbc{\ut{(P)}}
\def\Dbc{\ut{(D)}}
\def\Nbc{\ut{(N)}}

\def\izero{^{(0)}}
\def\ione{^{(1)}}
\def\itwo{^{(2)}}
\def\ithree{^{(3)}}
\def\ifour{^{(4)}}

\def\chit{\tilde\chi}

\def\d{\mathrm{d}}

\newcommand{\bitem}{\begin{itemize}}
\newcommand{\eitem}{\end{itemize}}
\newcommand{\benum}{\begin{enumerate}}
\newcommand{\eenum}{\end{enumerate}}
\newcommand{\btab}[1]{\begin{tabular}{#1}}
\newcommand{\etab}{\end{tabular}}
\newcommand{\beq}{\begin{equation}}
\newcommand{\eeq}{\end{equation}}
\newcommand{\beqn}{\begin{equation*}}
\newcommand{\eeqn}{\end{equation*}}

\newcommand{\bv}[1]{\mathbf{#1}}

\graphicspath{{Figs/}{figs/}}

\usepackage{bbold}
\newcommand{\cor}[1]{\mathcal{#1}}									
\newcommand{\T}[1]{\text{#1}}										
\newcommand{\eg}{\textit{e.g.}}
\newcommand{\ie}{\textit{i.e.}}
\newcommand{\n}{\nonumber}

\begin{document}
\title{Tracer particle in a confined correlated medium: an adiabatic elimination method}
\author{Davide Venturelli}
\email{davide.venturelli@sissa.it}
\affiliation{SISSA -- International School for Advanced Studies and INFN, via Bonomea 265, 34136 Trieste, Italy}
\author{Markus Gross}
\email{gross@is.mpg.de}
\affiliation{Max-Planck-Institut f\"{u}r Intelligente Systeme, Heisenbergstra{\ss}e 3, 70569 Stuttgart, Germany}
\affiliation{IV.\ Institut f\"{u}r Theoretische Physik, Universit\"{a}t Stuttgart, Pfaffenwaldring 57, 70569 Stuttgart, Germany}
\date{\today}

\begin{abstract}
We present a simple and systematic procedure to determine the effective dynamics of a Brownian particle coupled to a rapidly fluctuating correlated medium, modeled as a scalar Gaussian field, under spatial confinement. The method allows us, in particular, to address the case in which the fluctuations of the medium are suppressed in the vicinity of the particle, as described by a quadratic coupling in the underlying Hamiltonian. As a consequence of the confinement of the correlated medium, the resulting effective Fokker-Planck equation features spatially dependent drift and diffusion coefficients. We apply our method to simplified fluid models of binary mixtures and microemulsions near criticality containing a colloidal particle, and we analyze the corrections to the stationary distribution of the particle position and the diffusion coefficient.
\end{abstract}
\maketitle

\tableofcontents

\section{Introduction}
Determining effective equations of motion of a complex interacting system is a classic problem in Statistical Mechanics \cite{stratonovich_topics_1963, miguel1980colored, haenggi_colored_1995, gardiner_stochastic_2009, pavliotis_stochastic_2014}.
A typical scenario of this kind is encountered when a colloidal particle is immersed in a rapidly fluctuating medium. 
When there exists a clear separation between time scales, the fast degrees of freedom (medium) can be integrated out and subsumed into a reduced set of equations of motion for the slow degrees of freedom (particle).
Several schemes have been proposed in the past to achieve this goal, generally starting form a set of (stochastic) differential equations which are phenomenologically assumed to describe the coupled system composed by the particle and its bath.
Most of these schemes are based on the projection operator formalism \cite{mori_contraction_1980, morita_contraction_1980, zwanzig_non-equilibrium_2001, teVrugt_2020} or on the eigenfunction expansion of the Fokker-Planck equation corresponding to the original set of Langevin equations \cite{kaneko_adiabatic_1981, theiss_systematic_1985, theiss_remarks_1985, risken_fokker-planck_1989}.

The case in which the tracer particle moves in a correlated medium, which displays long-range correlations and large relaxation times, has recently received renewed attention \cite{demery_drag_2010, dean_diffusion_2011, demery_thermal_2011, demery_perturbative_2011, demery_diffusion_2013, fujitani_fluctuation_2016, Fujitani_2017, gross_dynamics_2021, Venturelli_2022, Basu_2022, Venturelli_2022_2parts}.
Indeed, this paradigm is relevant to describe, \eg, the dynamics of inclusions in lipid membranes \cite{reister_lateral_2005, reister-gottfried_diffusing_2010, camley_contributions_2012, camley_fluctuating_2014, Stumpf_2021}, microemulsions \cite{Gompper_1994,Hennes_1996,Gonnella_1997,Gonnella_1998}, as well as defects in ferromagnetic systems \cite{demery_drag_2010, demery_perturbative_2011}.
Moreover, recent advances in experimental technology have made it possible to measure the critical Casimir forces acting on colloidal particles immersed in near-critical fluid media \cite{hertlein_direct_2008,gambassi_critical_2009,magazzu_controlling_2019}. The properties of these forces, which are the classical thermal counterpart of the celebrated Casimir force in electromagnetism \cite{casimir_attraction_1948}, are relatively well understood in equilibrium \cite{krech_casimir_1994,kardar_friction_1999,brankov_theory_2000,gambassi_casimir_2009,maciolek_collective_2018,Dantchev_2022}, while much remains to be unveiled about their \textit{non-equilibrium} properties. Simple toy models have thus been devised in order to investigate the coupled out of equilibrium dynamics of one or a few particles in contact with near-critical media \cite{demery_drag_2010, demery_perturbative_2011, dean_diffusion_2011, zakine_spatial_2020, gross_dynamics_2021, Venturelli_2022, Basu_2022}. 
Theoretical descriptions of this type of systems often feature a scalar order parameter (OP) $\phi(\rv)$, which may represent the relative concentration of the two species in a binary liquid mixture, or else the deviation of the local fluid density from its critical value in a single-component fluid (\ie, $\phi(\rv)\propto n-n_c$). A velocity field $\vv (\rv,t)$ should be included to allow for the hydrodynamic transport of the OP field and the particle, but we will neglect it here for simplicity \cite{hohenberg_theory_1977}. A possible enhancement or suppression of the OP field or its correlations at the tracer location $\Rv(t)$ (representing, e.g., critical adsorption \cite{diehl_field-theoretical_1986}) can be modeled by including in the free energy terms proportional to $\phi(\Rv)$ or $\phi^2(\Rv)$, respectively, or derivatives thereof \cite{demery_thermal_2011}.

In order to obtain the effective dynamics of the tracer, a systematic method for the elimination of the field degrees of freedom coupled to $\Rv (t)$ from the dynamics is required -- particularly in the case in which the field relaxation is \textit{fast} (compared to the scale of the particle diffusion), but not instantaneous. The problem of the elimination of $N$ (non-interacting) fast variables from a system of $N+M$ stochastic differential equations is of course not new. 
The eigenfunction expansion method \cite{kaneko_adiabatic_1981,theiss_systematic_1985,theiss_remarks_1985} has been successfully applied to the case in which the field-particle coupling is linear, both in a confined geometry \cite{naji_hybrid_2009,gross_dynamics_2021} and in the continuum \cite{Venturelli_2022}. In this setting, the fast variables correspond to the Fourier modes of the field, while the $d$-dimensional coordinates of the particle are the slow variables.
The case of a quadratic coupling has hitherto not been addressed within the above formalism: notably, its underlying working hypothesis that the $N$ fast variables are non-interacting is generally violated, because the coupling to the particle can introduce interactions between the Fourier modes of the field. Other methods \cite{gardiner_adiabatic_1984, gardiner_stochastic_2009} require the specification of a steady-state distribution around which to construct a perturbative series in small powers of an appropriate \textit{adiabaticity} parameter $\chi$; however, such steady-state distribution is generally not known \textit{a priori} when the system is out of equilibrium.

In this paper we present an adiabatic elimination method for a tracer particle with generic (linear or quadratic) couplings to a stochastic background field in confinement. The method consists in a systematic expansion in powers of a small adiabaticity parameter $\chi$, which encodes the ratio of the field relaxation timescale to that of the particle diffusion. The procedure is transparent as it is based on a multiple-time-scale (or Chapman-Enskog-like) approach, where we project the dynamics over the \textit{moments} of the joint probability distribution $P(\Rv,\phi,t)$: this way we obtain a hierarchy of equations for the various moments, which can be truncated by noting that the higher moments relax faster than the lower ones \cite{Cates_2013,Solon_2015, Vishen_2018}.
Notably, no assumption is made \textit{a priori} about the equilibrium distribution of the tracer particle: the steady-state distribution is instead obtained from the method itself, the outcome of which is an effective Fokker-Planck (FP) equation for the tracer position $\Rv (t)$. In fact, our method can be applied to systems violating detailed balance between tracer and field degrees of freedom.

In the following, we will consider a point-like particle in contact with a scalar OP $\phi(\rv,t)$, whose Hamiltonian is considered within the Gaussian approximation, and which undergoes a Langevin relaxational dynamics (model A or B in the nomenclature of Ref.\ \cite{hohenberg_theory_1977}). We consider a box with periodic boundary conditions (BCs) for the field and the tracer in all but the $z$-direction, which has size $L$. In our actual calculations, we focus on a one-dimensional box of size $L$ in order to arrive at analytically tractable results. The field is spatially confined by suitable boundary conditions at $z=0,L$; a Brownian particle is allowed to diffuse in the box (subject to reflective \bcs at $z=0,L$), while coupled either linearly or quadratically to $\phi(\Rv,t)$, or to its derivatives $\nabla^n \phi(\Rv,t)$.
The backreaction of the particle on the evolution of the OP may or may not be taken into account. In the former case the tracer is termed \textit{reactive}, and it can be viewed as a model for a colloidal particle in a critical fluid; detailed balance is satisfied along its evolution, so that the system reaches equilibrium by relaxing to the Gibbs state. Conversely, in the latter case, a \textit{passive} tracer particle is carried by the medium without influencing it: it can be seen as an active particle driven by temporally correlated noise \cite{gross_dynamics_2021}. 
A main result of our study is an effective Fokker-Planck equation for the tracer position, characterized by space-dependent drift and diffusion coefficients.

The rest of the presentation is organized as follows. In \cref{sec_model} we introduce the model and set the notation. In \cref{sec_prelim} we recast the dynamics in terms of the field eigenfunctions expansion, and we exemplify some choices of free energies and field-particle couplings by pointing out their significance in physical applications. In \cref{sec_elimination} we develop the adiabatic elimination method, carrying out the derivation for a reactive or a passive tracer separately. As an example, in \cref{sec_applications} we apply the method to a few simple models, and we point out its qualitative predictions. Our results are finally summarized in \cref{sec_conclusions}.

\section{Model}
\label{sec_model}
We consider the joint dynamics of a point-like tracer particle at position $\Rv(t)$ coupled to a fluctuating order parameter scalar Gaussian field $\phi(\rv,t)$. The system is described by the Hamiltonian
\beq
\begin{split} 
    \Hcal(\Rv,[\phi]) \equiv \Hcal_\phi[\phi] + \Hcal_R(\phi(\Rv))&,\\
    \Hcal_\phi[\phi] \equiv  \int_V \d^d r\, \left\{ \onehalf \phi(\rv)\Delta(\rv)\phi(\rv)  - h_1 \phi(\rv)[ \delta(z) + \delta(L-z)] \right\},\qquad 
    &\Hcal_R(\phi(\Rv)) \equiv \frac{c}{2} (\cor{K}_2\phi(\Rv))^2 - h \cor{K}_1\phi(\Rv),
\end{split}\label{eq_Hamilt}
\eeq 
where we take $\cor{K}_1$, $\cor{K}_2$, $\Delta$ to be generic self-adjoint differential operators (see \cref{par:examples} below for examples), and the $\delta$-functions are assumed to be located inside the volume $V$. The system is confined in the $z$-direction, \ie, $z\in[0,L]$, and the OP field is required to fulfill one of the following boundary conditions (\bcs):
\begin{subequations}
\begin{align}
 \text{Dirichlet:}\qquad & \phi(\{\rvp,z\in\{0,L\}\})=0, \label{eq_bcs_Dir}\\
 \text{Neumann:} \qquad & \pd_z\phi(\rvp,z)|_{z\in\{0,L\}}=0, \label{eq_bcs_Neu}\\
 \text{capillary:} \qquad & \pd_z\bra\phi(\{\rvp,z\})\ket|_{z\in\{0,L\}}=\mp h_1 \label{eq_cap_bcs},
\end{align}\label{eq_bcs}
\end{subequations}
where $\rv=\{\rvp,z\}$. The terms proportional to the boundary fields $h_1$ in \cref{eq_Hamilt} induce capillary \bcs on the mean OP \cite{gross_dynamics_2021}.

Both the field and the particle are subject to a relaxational dynamics, as ruled by the Langevin equations \cite{dean_diffusion_2011,demery_perturbative_2011,demery_diffusion_2013}
\begin{subequations}
    \begin{align}
        \dot \Rv(t) &= -\gamma_R \nabla_\Rv \Hcal + \sqrt{\gamma_R} \bm{\eta}(t) , \label{eq:langevin_part}\\
        \dot\phi(\rv,t) &= -\gamma_\phi \Lambda \frac{\delta}{\delta \phi(\rv)} \left[ \Hcal_\phi[\phi] + \zeta \Hcal_R(\phi(\Rv)) \right] + \sqrt{\gamma_\phi} \xi(\rv,t), \label{eq:langevin_field}
    \end{align} \label{eq:langevin}
\end{subequations}
where $\Lambda(\rv)$ is another self-adjoint differential operator, and the noise terms represent two independent Gaussian stochastic processes with zero average and variances
\begin{align}
        \bra \eta_\alpha(t) \eta_\beta(t') \ket &= 2  T_R \delta_{\alpha\beta} \delta(t-t'),\\
        \bra \xi(\rv,t)\xi(\rv',t')\ket &= 2  T_\phi \Lambda(\rv) \delta(\rv-\rv')\delta(t-t').
        \label{eq:noise_field}
\end{align} 
The Fokker-Planck equation (FPE) associated with \cref{eq:langevin} is given by [$P\equiv P(\Rv,[\phi],t)$] \cite{gardiner_stochastic_2009}
\beq 
    \pd_t P = \int_V \d^d r \frac{\delta}{\delta \phi(\rv)} \left[\gamma_\phi \Lambda(\rv) \frac{\delta \Hcal}{\delta\phi(\rv)} + T_\phi \gamma_\phi \Lambda(\rv) \frac{\delta}{\delta\phi(\rv)}\right] P + \gamma_R \nabla_\Rv\cdot\left(\nabla_\Rv \Hcal \right)P + T_R \gamma_R \nabla^2_\Rv P,
    \label{eq_FPE_funct}
\eeq 
Equation \eqref{eq:langevin_part} describes the dynamics of the particle under the influence of the field $\phi(\Rv,t)$. The parameter $\zeta$ in \cref{eq:langevin_field} controls the back-reaction of the particle on the field: when $\zeta=0$ the field dynamics is independent from that of the particle, which we will call ``passive'' (in the sense that the particle is passively carried by the medium); for $\zeta=1$, instead, the field is influenced by the tracer, called henceforth ``reactive''. In the first case ($\zeta=0$), detailed balance is broken and the system is out of equilibrium for any choice of the temperatures $T_R$ and $T_\phi$; for all practical purposes, \cref{eq:langevin_part} then describes an active particle driven by the (independent) stochastic process in \cref{eq:langevin_field}. Within the second scenario ($\zeta=1$), on the other hand, setting $T_R=T_\phi\equiv T$ corresponds to a situation in which the particle and the field are in contact with the same thermal bath; the resulting steady-state probability distribution [which solves \cref{eq_FPE_funct}] is then given by the Gibbs state
\beq 
    P_s(\Rv,[\phi]) = \frac{1}{\Zcal_0} \exp\left(-\frac{1}{T}\Hcal(\Rv,[\phi]) \right),
    \label{eq_Pss_joint}
\eeq
with $\Zcal_0 = \int_V\d^{d} R \int \Dcal\phi\, \exp(-\Hcal(\Rv,[\phi])/T)$, and the functional measure $\cor{D} \phi$ is as usual defined as
\beq
    \int \cor{D}\phi \equiv \prod_n \int_{-\infty}^\infty \d \phi_n 
    \label{eq:path_measure}
\eeq
in terms of suitable field eigenmodes $\phi_n$ -- see \cref{sec:mode_expansion}.

In the following, we will focus for simplicity on one spatial dimension, and we will be interested in the marginal dynamics of the tracer particle $\Rv(t)$ alone. Under the assumption of fast field relaxation (a notion we will make more precise in \cref{sec_dynamics} below), we will write an approximate effective Fokker-Planck equation for the marginal probability distribution 
\beq \bar P(\Rv,t) \equiv \int\Dcal\phi \, P(\Rv,[\phi],t) \, ,
\label{eq_P_margin}\eeq
which, due to global conservation of probability, fulfills
\beq \int_V \d^d R \, \bar P(\Rv,t) = 1 \, .
\label{eq_P_cons}\eeq 
We choose for simplicity the initial joint probability distribution to be given by
\beq P(\Rv,[\phi], t= t_0) = \delta(\Rv-\Rv_0) P_\phi(\Rv_0,[\phi]) \, .
\label{eq_P_initcond}\eeq 
We assume the initial condition $\phi(\rv,t=t_i)=0$ to apply in the infinite past ($t_i = -\infty$), so that at $t=t_0=0$ the OP has reached a steady state and its initial condition can be neglected. The initial condition for the particle is applied at $t=t_0=0$.

\section{Preliminaries}
\label{sec_prelim}

\subsection{Mode expansion of the OP field}
\label{sec:mode_expansion}
Unless stated otherwise, we focus henceforth on a one-dimensional system with $z\in[0,L]$.
It is convenient to expand the field variable $\phi(r,t)$ in terms of a suitable eigenfunction basis.
Assuming that the operators $\Delta(r)$ and $\Lambda(r)$ commute, we can find a set of common eigenfunctions $\sigma_n$ satisfying the eigenvalue equations
\begin{equation}
    \Delta \sigma_n \equiv \beta_n \sigma_n \, , \qquad
    \Lambda \sigma_n \equiv \frac{1}{T_\phi}L_n \sigma_n \, ,
\label{eq_eigenval}\end{equation}
where $n$ denotes the mode index [see \cref{eq_eigenspec} below for specific expressions].
We do not require that $\Delta(r)$, $\Lambda(r)$ commute with $\cor{K}_{1,2}(r)$; we thus define
\begin{equation}
    u_n(z) \equiv \sqrt{c} \cor{K}_2\sigma_n(z) \, , \qquad
    v_n(z) \equiv h \cor{K}_1 \sigma_n(z) \, , 
    \label{eq:u_and_v}
\end{equation}
where $\sigma_n$ is not in general an eigenfunction of $\cor{K}_2$ or $\cor{K}_1$.

We then introduce the following eigenfunction expansions of the OP and the noise fields:
\begin{equation}
    \phi(z,t) = \sum_{n} \sigma_n(z) \phi_n(t),\qquad 
    \xi(z,t) = \sum_{n} \sigma_n(z) \xi_n(t) ,
    \label{eq_phi_expand}
\end{equation}
where the expansion coefficients are defined through the inverse relations
\begin{equation}
    \phi_n(t) = \int_0^L\d z\, \sigma^*_n(z) \phi(z,t),\qquad
    \xi_n(t) = \int_0^L\d z\, \sigma^*_n(z) \xi(z,t).
    \label{eq:fourier_coefficients}
\end{equation}
The eigenfunctions $\sigma_n(z)$ are taken to be orthonormal,
\beq 
    \int_0^L \d z\, \sigma_m^*(z) \sigma_n(z) = \delta_{m,n},
    \label{eq_eigenf_ortho}
\eeq 
and they satisfy the completeness relation
\beq 
    \sum_n \sigma_n(z) \sigma_n^*(z') = \delta(z-z').
    \label{eq_eigenf_complete}
\eeq
Unless otherwise noted, we will consider in the following only real eigenfunctions $\sigma_n$ (see also the discussion in \cref{sec_eigenbasis}).

\subsection{Stationary distribution and effective potentials in the presence of detailed balance}
\label{sec_stationary_potentials}
Let us first introduce the notation
\begin{align}
    \Gamma_{nm} &\equiv \frac{\beta_n}{T_\phi}\delta_{nm} + \frac{1}{T_R}u_n(R) u_m(R)\, , \label{eq:Gamma} \\
    \tau_n &\equiv  \frac{h_1}{T_\phi}[\sigma_n(0)+\sigma_n(L)] +  \frac{\zeta}{T_R} v_n(R) \, , \label{eq:tau}
\end{align}
where $\beta_n$, $u_n$ and $v_n$ are defined as in \cref{sec:mode_expansion}.

As we stressed above, the evolution equations \eqref{eq:langevin} satisfy detailed balance in the reactive case ($\zeta=1$) and provided that $T_R=T_\phi\equiv T$. In this case, the joint stationary distribution is the one given in \cref{eq_Pss_joint}, and it can be expressed in terms of the OP modes as
\begin{equation}
    P_s(R,\{ \phi_n \}) = \frac{1}{\cor{Z}_0}\exp\left\lbrace-\frac12 \sum_{nm} \phi_n \Gamma_{nm} \phi_m  +\sum_n \tau_n \phi_n \right\rbrace \, .
    \label{eq:stationary}
\end{equation}
The marginal stationary probability distribution of the particle at position $R$ can be found by integrating out the field modes $\phi_n$ from the joint stationary distribution $P_s(R,\{ \phi_n \})$ in \cref{eq:stationary}. The latter is Gaussian in the $\phi_n$'s, so that a straightforward calculation renders (when $T_R=T_\phi\equiv T$)
\beq
    P_s(R) \equiv \int \cor{D}  \phi \, P_s(R,\{ \phi_n \}) = \cor{N} \exp\left\lbrace -\frac{1}{T}[ U(R) + W(R)] \right\rbrace \, ,
    \label{eq:marginal_stationary}
\eeq
where $\cor{N}$ is a normalization constant, and where we introduced the effective potentials
\begin{align}
    U(R) &\equiv \frac{T}{2} \ln(1+V(R)) \, , \label{eq:U_(R)}\\
    V(R) &\equiv \sum_n \frac{u_n^2(R)}{\beta_n} \, , \label{eq:V(R)} \\
    W(R) &\equiv -\frac{T}{2} \tau_n \Gamma_{nm}^{-1} \tau_m \, , \label{eq:W(R)}
\end{align}
stemming from the use of the matrix determinant lemma \cite{Matrix_2012} [in partricular, $U(R)$ follows from the determinant of $\Gamma_{nm}$ in \cref{eq:Gamma}].
Note that $W(R)$ contains the effect of possible boundary fields $h_1 \neq 0$ through $\tau_n$ defined in \cref{eq:tau}, while the inverse matrix $\Gamma^{-1}$ can be explicitly computed by means of the Shermann-Morrison formula \cite{Matrix_2012}, yielding
\beq
    \Gamma_{ij}^{-1} = T \left[ \beta_i^{-1} \delta_{ij} -  \frac{u_i u_j}{\beta_i \beta_j (1+V(R))} \right] \, .
    \label{eq:gamma_inverse}
\eeq
The effective potentials $V(R)$ and $W(R)$ are model-dependent, and they involve an infinite summation of terms containing the linear or quadratic coupling to the particle in their numerator (i.e., the functions $v_n(R)$ or $u_n(R)$, respectively), and the eigenvalue $\beta_n$ of the field operator $\Delta(r)$ in their denominator. These sums can occasionally diverge, either because the denominator of the $n=0$ term is zero [called infrared (IR) divergence], or because the series does not converge to a finite result at large $n$ [ultraviolet (UV) divergence]. IR divergences are typically related to the presence of zero modes in the OP field at criticality; we will see some examples of them in \cref{sec_applications}, and we will comment later on their meaning. In contrast, UV divergences are more subtle: they may either result in a flat stationary probability distribution (which is physically meaningful), or else prevent the said probability density function (PDF) from being normalized \cite{gross_dynamics_2021}. This second type of divergence generally indicates that the corresponding effective potential is \textit{non-universal}, in the sense that it depends on other UV details which were not included in the original Hamiltonian. Indeed, these divergences are generally cured by including higher-order derivative terms in the field Hamiltonian $\cor{H}_\phi$ in \cref{eq_Hamilt}, which in turn translates into the corresponding eigenvalue $\beta_n$ being proportional to higher powers of the summation index $n$. 

In \cref{sec_applications} we will exemplify the use of the adiabatic approximation procedure by applying it to models where the effective potentials in \cref{eq:U_(R),eq:V(R),eq:W(R)} are not affected by UV divergences. 

\subsection{Some concrete examples}
\label{par:examples}
Many physical models fall within the class of Hamiltonians introduced in \cref{eq_Hamilt} when they are considered within their quadratic (Gaussian) approximation. The simplest is arguably the Landau-Ginzburg model endowed with model A/B dynamics \cite{hohenberg_theory_1977}, which corresponds to the choice of operators [see \cref{eq_eigenval}]
\begin{align}
    \Delta(\rv) &= -\nabla_\rv^2+\tau \qquad \to \qquad
    \beta_n = k_n^2+\tau \, , \label{eq:Delta_example} \\
    \Lambda(\rv) &= (-\nabla_\rv^2)^a \qquad \to \qquad L_n = T_\phi k_n^{2a} \, , \label{eq:Lambda_example}
\end{align}
where the parameter $a$ can take the value $a=0$ (model A, dissipative dynamics) or $a=1$ (model B, conserved dynamics).
For the sake of generality, we consider here again arbitrary dimension $d$ and, accordingly, the squared wave number $k_n^2$ stands for $\sum_{i=1}^d k_{i,n}^2$ [see \cref{eq_eigenspec} below for expressions of $k_{z,n}$ in the one-dimensional case].
The parameter $\tau$ can be linked to the correlation length $\xi$ of the field by $\tau=\xi^{-2}$, so that for $\tau\to 0$ the system approaches a critical point characterized by the divergence of $\xi$ \cite{krech_casimir_1994, brankov_theory_2000, maciolek_collective_2018,Dantchev_2022}.

Membranes are often described in terms of the Helfrich Hamiltonian \cite{Helfrich_1973,Campelo_2014,Stumpf_2021,seifert1997configurations}, for which
\begin{equation}
    \Delta(\rv) = \kappa \nabla_\rv^4 - \sigma \nabla_\rv^2 \qquad \to \qquad  
    \beta_n = \kappa k_n^2(k_n^2+\tau) \, ,
\end{equation}
where $\kappa$ is the membrane bending modulus and $\sigma$ its surface tension, and where we introduced $\tau=\sigma/\kappa$. 

Microemulsions (\eg, oil-water-surfactant mixtures) can instead be described in terms of the Gompper-Schick model \cite{Gompper_1994,Hennes_1996}, where to the scalar order parameter $\phi(\rv,t)$ one generically associates a free energy
\begin{equation}
    \cor{F}[\phi] = f(\phi) + c_2(\nabla \phi)^2 + c_4 (\nabla^2 \phi)^2 \, ,
    \label{eq:free_energy_GS}
\end{equation}
where $f(\phi)$ is a local polynomial in $\phi$ (typically of the 4th or 6th order). Choosing again model A/B dynamics leads to a 4th/6th order Allen-Cahn/Cahn-Hiliard evolution equation for the order parameter $\phi(\rv,t)$, a problem which has been extensively covered in the mathematical literature \cite{Paetzold_1995,Pawlow_2011,Schimperna_2011, Yang_2018}. When not specifically interested in studying phase coexistence, one can consider this model within the Gaussian approximation by truncating $f(\phi)\simeq \tau \phi^2$ to the quadratic order, again leading to
\begin{equation}
    \Delta(\rv) = c_4 \nabla_\rv^4 - c_2 \nabla_\rv^2 +\tau \qquad \to \qquad
    \beta_n = c_4 k_n^4 +c_2 k_n^2+\tau \, .
    \label{eq:operators_GS}
\end{equation}
In contrast to the Helfrich model, in the GS model the constant $c_2$ can take both positive or negative values: this can be used to describe the enhancement/reduction of the surface tension by the surfactant.

While the class of Hamiltonians in the form of \cref{eq_Hamilt} is broad, 
finding a simultaneous eigenbasis for the operators $\Lambda(\rv)$ and $\Delta(\rv)$ is not always straightforward (albeit unfortunately necessary in order to obtain closed-form results for the effective Fokker-Planck equation, as we will discuss later).
For instance, the Helfrich Hamiltonian is often used in conjunction with a dynamics ruled by the Oseen hydrodynamic tensor \cite{seifert1997configurations, Naji_2007,camley_fluctuating_2014}. In $d=2$ and in Fourier space, the Oseen tensor takes the approximate form $\Lambda(\qv) = 1/(4\eta |\qv|)$, where $\eta$ is the viscosity of the fluid surrounding the membrane \cite{Brown_2004, demery_diffusion_2013}. While in the bulk one can use plane waves $\exp(i \qv\cdot \rv)$ to diagonalize $\Lambda(\rv)$ and $\Delta(\rv)$ simultaneously, in a confined geometry such as the one considered in this work, the Fourier series coefficients of $\Lambda(\qv) = 1/(4\eta |\qv|)$ assume a cumbersome form which makes the calculation less practical. We will thus reserve the analysis of models involving the Oseen tensor for a future study.

Finally, a tracer particle can be coupled to the order parameter by means of the linear and/or quadratic terms contained in $\cor{H}_R$ in \cref{eq_Hamilt}. Typical choices for the operators $\cor{K}_1$ and $\cor{K}_2$ are the identity $\mathbb{1}$, or else powers of $\nabla_\Rv$; we will sometimes denote the former case as \textit{simple} linear or quadratic couplings, \ie,
\begin{equation}
    \Kcal_1=\mathbb{1},\quad \Kcal_2=\mathbb{1},\quad  \Hcal_R(\phi(\Rv)) \equiv \frac{c}{2} \phi^2(\Rv) - h \phi(\Rv) \, .
    \label{eq:simple_couplings}
\end{equation}
In the linear coupling case (\eg, $h\phi(\Rv)$ or $h \nabla_\Rv \phi(\Rv)$) and for $h>0$, configurations are favored in which the OP field (or its derivative) are enhanced in the vicinity of the tracer particle. In the quadratic coupling case (\eg, $c\phi^2(\Rv)$ or $c [\nabla_\Rv \phi(\Rv)]^2$), on the contrary, the value of the OP field (or its derivative, respectively) are suppressed in the vicinity of the tracer particle; in the formal limit where $c\to \infty$, the coupling induces a point-like Dirichlet (or Neumann) boundary condition on $\phi$ in correspondence of the tracer's position $\Rv$ \cite{demery_thermal_2011}.

\subsection{Choice of the eigenbasis}
\label{sec_eigenbasis}
Local differential operators $\Lambda(z)$, $\Delta(z)$ like the ones listed above are all diagonalized by the following eigenfunctions $\sigma_n(z)$, which we specialize for the various \bcs considered in \cref{eq_bcs}:
\begin{subequations}
\begin{align}
\sigma_n\Dbc(z) &= \sqrt{\frac{2}{L}} \sin\left(k_n\Dbc z\right), \qquad k_n\Dbc = \frac{\pi n}{L}, \qquad n=1,2,\ldots,  \qquad &\text{Dirichlet \bcs}, \label{eq_eigenf_Dbc}\\
\sigma_n\Nbc(z) &= \sqrt{\frac{2-\delta_{n,0}}{L}} \cos\left(k_n\Nbc z\right), \qquad k_n\Nbc = \frac{\pi n}{L}, \qquad n=0,1,2,\ldots,  \qquad &\text{Neumann \bcs}. \label{eq_eigenf_Nbc}
\end{align}\label{eq_eigenspec}
\end{subequations}
Capillary \bcs [see \cref{eq_cap_bcs}] can be imposed by choosing Neumann eigenfunctions, for which the averaged OP profile can be shown to read \cite{gross_dynamics_2021}
\beq 
    \bra\phi(z)\ket_{h_1}   =  h_1 L \left[\left( \frac{1}{2} - \frac{z}{L} \right)^2 - \frac{1}{12}\right].
    \label{eq_avg_prof_h1}
\eeq 
We furthermore introduce, for future reference, the shorthand notation
\beq \tilde\sigma_n(z) \equiv \frac{1}{k_n} \pd_z\sigma_n(z) = \begin{cases}
                          \sqrt{\frac{2}{L}}\cos(k_n\Dbc z) ,\qquad n=1,2,\ldots &\qquad \text{(D)}\\
                          -\sqrt{\frac{2}{L}}\sin(k_n\Nbc z) ,\qquad n=1,2,\ldots &\qquad \text{(N)} 
                         \end{cases}
\label{eq_eigenf_deriv}\eeq
and $\tilde\sigma_0\Nbc(z) = 0$.

It must be noted that standard Dirichlet BCs generally entail a non-zero flux through the boundaries \cite{diehl_boundary_1992, gross_dynamics_2019}, requiring to use suitable ‘no-flux’ basis functions in order to ensure global OP conservation \cite{gross_first-passage_2018-1}. This is however technically rather involved, so in the following we will not consider conserved dynamics (model B) in conjunction with Dirichlet BCs.

In this work we are mostly concerned with the effects of confinement on the dynamics of the tracer particle, and thus we will not focus explicitly on the case of periodic boundary conditions (PBCs), for which most of these effects have been found to trivialize \cite{gross_dynamics_2021}. However, the choice of eigenfunctions
\beq 
    \sigma_n\Pbc(z) \equiv  \begin{cases}
                          \frac{1}{\sqrt{2}}\sigma_n\Nbc (z) ,& n=0,-1,-2,\ldots \\
                          \frac{1}{\sqrt{2}}\sigma_n\Dbc (z) ,& n=1,2,3,\ldots
                         \end{cases}
    \label{eq_eigenf_periodic_real}
\eeq
formally corresponds to PBCs over the symmetric interval $z\in[-L,L]$ (which is a convenient choice for later inspecting the \textit{bulk} limit $L\to\infty$).
These could be equivalently addressed by adopting complex eigenfunctions $\sigma_n\pbc(z) = \frac{1}{\sqrt{2L}} \exp(\im k_n\pbc z) , k_n\pbc = \pi n/L$, but then the FP-based method we present in \cref{sec_elimination} would require considering the joint probability distribution $\cor{P}(\alpha_R,\alpha_I,t)$ [or equivalently $\cor{P}(\alpha,\bar \alpha,t)$, with $\bar \alpha \equiv \alpha^*$] for any complex dynamical variable $\alpha=\alpha_R + i \alpha_I$. While this is technically straightforward, the number of dynamical operators would double in our entire derivation. In order to limit the proliferation of terms and for the sake of clarity, we will assume in the following that the eigenfunctions $\sigma_n$ are chosen real.

\subsection{Dynamics}
\label{sec_dynamics}
Let us rescale time as $t \to t/\gamma_R$ in the Langevin equations \eqref{eq:langevin}, and define appropriately rescaled fields (cf.\ \cite{gross_dynamics_2021}); this is equivalent to setting $\gamma_R\equiv 1$ in \cref{eq:langevin_part} and replacing $\gamma_\phi \to \chi^{-1}$ in \cref{eq:langevin_field}, where we introduced the ``adiabaticity'' parameter
\beq 
    \chi\equiv \sfrac{\gamma_R}{\gamma_\phi} \, .
    \label{eq_chi}
\eeq 
A dimensionless counterpart of $\chi$ in \cref{eq_chi} can be introduced as follows: 
\beq 
    \chit \equiv T_R L^{-d_\Lambda} \chi \, ,
    \label{eq_adiab_param}
\eeq 
where we denoted by $d_\Lambda \equiv [\Lambda(r)]$ the length dimension of the dynamical operator $\Lambda(r)$ introduced in \cref{eq:langevin_field}, and we have furthermore chosen $T_R$ in order to account for the temperature dimension of $\chi$.
In the \emph{adiabatic limit} $\chit\ll 1$, the OP dynamics is much faster than the tracer dynamics.

The coupled Langevin equations \eqref{eq:langevin_part} and \eqref{eq:langevin_field} can now be rewritten in terms of the eigenfunctions $\sigma_n$ as
\begin{align}
    \pd_t R &=- \sum_{nm} A_{nm} \phi_n\phi_m + \sum_n t_n \phi_n + \eta , \label{eqq_langevin_modes_R} \\
    \pd_t \phi_n &=- \chi^{-1} \left[ \sum_{m} B_{nm} \phi_m - s_n \right] + \chi^{-1/2}\xi_n \, .
    \label{eqq_langevin_modes}
\end{align}
Above we have introduced the vectors 
\begin{align}
    t_n &\equiv  \partial_R v_n(R) \, , \label{eq_tn_def}\\
    s_n& \equiv L_n \tau_n \, ,
\label{eq_sn_def}\end{align}
and the matrices
\begin{align}
    A_{ij} &\equiv \frac{T_R}{2}\partial_R \Gamma_{ij}  \equiv \frac12 \pd_R (u_i u_j)   \, , \label{eq:def_A} \\
    B_{ij} &\equiv  b_i \delta_{ij} + c_{ij} \, , \label{eq:def_B}
\end{align}
where we denoted
\begin{equation}
    b_i\equiv L_i\beta_i /T_\phi \, , \qquad c_{ij} \equiv \zeta L_i u_i u_j / T_R \, ,
    \label{eq:b_and_c}
\end{equation}
with $\beta_i$, $L_i$, $\Gamma_{ij}$ and $\tau_n$ defined in \cref{eq_eigenval,eq:Gamma,eq:tau}, respectively (see \cref{par:examples} for specialization to the various models).

The noise variances in \cref{eq:noise_field} can also be expressed in terms of its expansion coefficients $\xi_n(t)$ as
\beq 
    \bra \xi_m(t) \xi_n(t')\ket = 2  L_n \delta_{mn} \delta(t-t').
    \label{eq_phi_noise_mode_correl}
\eeq 
Equations \eqref{eqq_langevin_modes} and \eqref{eqq_langevin_modes_R} correspond to the following FPE for the joint probability distribution $P(R,\{ \phi_n \},t)$ :
\beq 
    \pd_t P = \pd_R \left[ \sum_{mn}  A_{nm} \phi_n\phi_m - \sum_n t_n \phi_n\right] P + T_R\pd_R^2 P +  \chi^{-1} \sum_{n} \pd_{\phi_n}\left[ \sum_m B_{nm} \phi_m - s_n\right] P + \chi^{-1} \sum_n L_n \pd_{\phi_n}^2 P \, .
    \label{eq:FPE_modes}
\eeq 
Finally, note that in the reactive case ($\zeta=1$) it follows from \cref{eq:Gamma,eq:tau} that ($L_{ij}\equiv L_i \delta_{ij}$)
\begin{align}
   B_{ij} &= (L \circ \Gamma)_{ij}\, , \label{eq:def_B_reactive} \\
    t_n &= T_R \pd_R \tau_n \label{eq:def_tn_reactive} \, .
\end{align}

\section{Adiabatic elimination method}
\label{sec_elimination}
In this Section we describe how the field coordinates $\phi_n$ can be eliminated from \cref{eq:FPE_modes}, thus yielding a Fokker-Planck equation which governs the dynamics of $R(t)$ alone, under the assumption that the field equilibrates faster than the particle. To this end, we define the following moments of $P(R,\{\phi_i\})$:
\begin{subequations}
\begin{align}
    Q\izero(R,t) &= \int \Dcal\phi\, P(R,\{\phi_i\}) \, , \\
    Q\ione_n(R,t) &= \int \Dcal\phi\, \phi_n(t) P(R,\{\phi_i \}) \, ,  \\
    Q\itwo_{nm}(R,t) &= \int \Dcal\phi\, \phi_n(t) \phi_m(t) P(R,\{ \phi_i\}) \, , \\
    Q\ithree_{nmp}(R,t) &= \ldots \, , 
\end{align}\label{eq_Q_mom}
\end{subequations}
where $\int\Dcal \phi $ indicates a multidimensional integral over the modes of $\phi$ [see \cref{eq:path_measure}].
Note that $Q\izero(R,t)=\bar P(R,t)$ [see \cref{eq_P_margin}] is the time-dependent marginal probability distribution of the particle position, whose dynamics we are interested in.
Integrating \cref{eq:FPE_modes} over a mode $\phi_n$ as in \cref{eq_Q_mom} provides a hierarchy of equations for the evolution of these moments:
\begin{subequations}
\begin{align}
    \pd_t Q\izero =& \pd_R \sum_{nm} A_{nm} Q\itwo_{nm}  -\pd_R \sum_{n} t_n Q_n\ione + T_R\pd_R^2 Q\izero \, , \label{eq:Q0} \\
    \pd_t Q_n\ione =&  \pd_R \sum_{ij} A_{ij} Q\ithree_{ijn} - \pd _R \sum_{m} t_m Q\itwo_{nm} + T_R \pd_R^2  Q_n\ione - \chi^{-1} \left[ \sum_{m} B_{nm} Q\ione_m - s_n Q\izero\right] \, , \label{eq:Q1} \\
    \pd_t  Q_{nm}\itwo =& \pd_R \sum_{ij} A_{ij} Q\ifour_{ijnm} - \pd _R \sum_{i} t_i Q\ithree_{inm} + T_R \pd_R^2 Q_{nm}\itwo \n\\
    &- \chi^{-1} \sum_{j}\left[B_{nj} Q\itwo_{jm} + B_{mj} Q\itwo_{jn} \right] + \chi^{-1} \left[ s_n Q_m\ione +s_m Q_n\ione \right] + 2\chi^{-1} L_n \delta_{nm} Q\izero \, , 
    \label{eq:Q2}
\end{align} \label{eq:adiabatic}
\end{subequations}
and so on. By working in the adiabatic limit $\chi \ll 1$, we can formally expand the moments in powers of small $\chi$:
\beq 
    Q^{(j)}_{nm\ldots} = q^{(j)}_{nm\ldots} + \chi \tilde q^{(j)}_{nm\ldots} + \Ocal(\chi^2) \, .
    \label{eq:Q_chi_exp}
\eeq 
The hierarchy of equations \eqref{eq:adiabatic} can then be closed by replacing the moments $Q^{(j)}_{nm\ldots}$ by their expansion in \cref{eq:Q_chi_exp}, and then neglecting terms of $\cor{O}(\chi^2)$ or higher. In the next Sections, we will thus solve \cref{eq:Q1,eq:Q2}, and plug the result back into \cref{eq:Q0} in order to obtain an evolution equation for the marginal probability distribution $Q\izero(R,t)$, \ie,
\beq 
    \pd_tQ\izero = -\pd_R [\mu(R) Q\izero(R)] + \pd_R^2 [D(R) Q\izero(R)] +\cor{O}(\chi^2) \, ,
    \label{eq:effective_FP}
\eeq
where $\mu(R)$ and $D(R)$ represent the space-dependent effective drift and diffusion coefficients, respectively.

In the following, we will consider separately the case of a reactive ($\zeta=1$) and passive ($\zeta=0$) tracer particle.

\subsection{Reactive case}
\label{par:reactive}
The lowest order in the adiabatic approximation is obtained by truncating the expansion in \cref{eq:Q_chi_exp} to its leading order ($\chi=0$); equivalently, we formally take the limit $\gamma_\phi\to \infty$ of the mobility of the field [see \cref{eq_chi}]. The physical meaning of this is that the field instantaneously rearranges around the position $R(t)$ assumed at each time $t$ by the particle -- this is reminiscent of the Born-Oppenheimer approximation in condensed matter physics \cite{Bransden_2014}. In the reactive case, setting $T_\phi=T_R=T$ implies that the steady-state probability distribution is given by $P_s(R,\{ \phi_i \})$ in \cref{eq:stationary}, because of detailed balance: accordingly, we can write
\beq
    P(R,\{ \phi_i \},t) = P_s(R(t),\{ \phi_i \}) +\cor{O}(\chi) \, .
    \label{eq:quasi_static}
\eeq
We thus deduce that we can obtain the lowest order terms $q^{(j)}_{nm\ldots}$ in the expansion of $Q^{(j)}_{nm\ldots}$ in \cref{eq:Q_chi_exp} by taking the Gaussian expectation values
\beq 
    q^{(j)}_{nm\ldots} = \int \Dcal\phi\, \phi_n \phi_m \ldots P_s(R(t),\{ \phi_i\}) \equiv \left\langle \phi_n \phi_m \ldots \right\rangle \, .
    \label{eq:expvals} 
\eeq
These are easily obtained by first constructing the generating functional \cite{gross_dynamics_2021}
\beq
    \cor{Z}[\{ J\}] = \int \Dcal\phi\, P_s(R(t),\{ \phi\}) \exp\left[ \sum_n (J_n +\tau_n) \phi_n \right] = \cor{Z}[0] \exp\left[ \frac12 \sum_{nm} J_n \Gamma_{nm}^{-1} J_m + \sum_{nm}  J_n \Gamma_{nm}^{-1} \tau_m \right] \, ,
\eeq
where $J(z) = \sum_n \sigma_n(z) J_n$ is an auxiliary field, while $\Gamma_{nm}$ and $\tau_n$ are defined in \cref{eq:Gamma,eq:tau}, respectively. Note that $\cor{Z}[0]=P_s(R)$, \ie, the marginal stationary distribution given in \cref{eq:marginal_stationary}. By Wick's theorem, one can obtain the joint cumulants
\begin{align}
    Q\izero &= \cor{Z}[0] \, , \n\\
    q_{i}\ione  &=  \cor{Z}[0] \Gamma_{ij}^{-1} \tau_j \equiv \cor{Z}[0] \bra i \ket \, ,  \n \\
    q_{ij}\itwo &= \cor{Z}[0] \left( \Gamma_{ij}^{-1} +  \Gamma_{im}^{-1} \Gamma_{jn}^{-1} \tau_m \tau_n \right) \equiv \cor{Z}[0] \bra ij \ket  \equiv \cor{Z}[0] \left[ \bra ij \ket_c + \bra i \ket \bra j \ket \right] \, ,  \n\\
    q_{ijk}\ithree &\equiv \cor{Z}[0] \bra ijk \ket = \cor{Z}[0] \left[ \bra i\ket \bra j\ket \bra k \ket + \bra ij \ket_c \bra k \ket + \T{(2 perm.)} \right] \, , \n\\
    q_{ijkl}\ifour &\equiv \cor{Z}[0] \bra ijkl \ket = \cor{Z}[0] \left[ \bra i\ket \bra j\ket \bra k \ket \bra l \ket  + \bra ij \ket_c \bra kl \ket_c + \T{(2 perm.)}  + \bra ij \ket_c \bra k \ket \bra l \ket + \T{(5 perm.)} \right] \, , 
    \label{eq:wickthm}
\end{align}
where we have introduced the shorthand notation $ \bra ij \ket_c \equiv \Gamma_{ij}^{-1} $ and $\bra i \ket \equiv \Gamma_{ij}^{-1} \tau_j $.

In this notation, the Langevin equation \eqref{eq:Q0} reads at lowest order
\begin{equation}
    \pd_t Q\izero = \pd_R \left[ U'(R) + W'(R) \right] Q\izero + T \pd^2_R Q\izero + \Ocal(\chi)  \, ,
    \label{eq:super_adiabatic}
\end{equation}
where the prime denotes a derivative with respect to $R$. Setting $\pd_t Q\izero \equiv 0$ promptly yields the correct marginal stationary distribution given in \cref{eq:marginal_stationary}. We will call this the \textit{super-adiabatic} approximation for the effective dynamics of the particle. Comparing \cref{eq:super_adiabatic} with \cref{eq:effective_FP} shows that, at the lowest order in the adiabatic approximation, the diffusion term is not modified, \ie, $D(R)=T$, while the drift term $\mu(R)=-\left[ U'(R) + W'(R) \right]$ is the one intuitively expected for a particle moving in the OP-induced effective stationary potential given in \cref{eq:marginal_stationary}. Although the latter seems to diverge for large $c$ (see \cref{eq:U_(R),eq:V(R),eq:u_and_v}), in fact its derivative does not, leading in the purely quadratic case ($h=h_1=0$) to an effective drift term 
\begin{equation}
    \mu(R) \xrightarrow[c\to \infty]{} -\frac{TV'(R)}{2V(R)} \, ,
\end{equation}
with $V(R)$ given in \cref{eq:V(R)}.

We remark that it is possible to obtain \cref{eq:super_adiabatic} without resorting to the \textit{quasi-static} assumption we made in \cref{eq:quasi_static}, but instead by directly solving a Lyapunov matrix equation (see \cref{app:lyapunov}). However, the latter route turns out to be computationally challenging when one tries to move beyond the $\cor{O}(\chi^0)$ result.

The $\Ocal(\chi)$ correction in \cref{eq:super_adiabatic} is obtained by computing $\tilde q\ione$ and $\tilde q\itwo$ from \cref{eq:adiabatic}. Using \cref{eq:def_A} and the matrix relation \cite{bellman_1997, Matrix_2012}
\beq
    \pd_R \cor{M}^{-1}(R) = -\cor{M}^{-1}\circ \pd_R \cor{M} \circ \cor{M}^{-1} \qquad \to \qquad \pd_R \Gamma^{-1} = -\frac{2}{T}  \Gamma^{-1}\circ A \circ \Gamma^{-1} \, ,
\eeq
a long but straightforward calculation renders
\begin{subequations}
\begin{align}
    B_{nm} \tilde q\ione_m &= \left( \pd_R \bra n \ket \right) \left[ (U' + W') Q\izero + T \pd_R Q\izero \right] \, , \label{eq:q1tilde} \\
    B_{nj} \tilde q\itwo_{jm} + B_{mj} \tilde q\itwo_{jn} &= s_n \tilde q\ione_m + s_m \tilde q\ione_n + \left( \pd_R \bra nm \ket \right) \left[ (U' + W') Q\izero + T \pd_R Q\izero \right] \, . \label{eq:q2tilde}
\end{align}
\label{eq:final_moments}
\end{subequations}
The terms $\tilde q\ione$ and $\tilde q\itwo$ in general provide corrections both to the drift $\mu(R)$ and to the diffusion coefficient $D(R)$ in \cref{eq:effective_FP}.
Note that these contributions vanish when using for $Q\izero$ the stationary distribution in \cref{eq:stationary}, thus guaranteeing that the effective FP equation will still satisfy detailed balance.

In the next subsections we will solve \cref{eq:q1tilde,eq:q2tilde} in the cases where either a linear or a quadratic coupling to the particle is included in the Hamiltonian in \cref{eq_Hamilt}. We report in \cref{app:mixed_case} the most general case in which both couplings are included. The consistency of our results with those obtained in the bulk limit $L\to\infty$ in Refs.~\cite{dean_diffusion_2011,demery_perturbative_2011,demery_diffusion_2013} will finally be explored in \cref{app:bulk}.

\subsubsection{Linear coupling}
In this case, we set $c=0$: this simplifies the calculation significantly, because then the matrices $B_{nm} = b_n \delta_{nm}= \delta_{nm} L_n \beta_n/T$ and $\Gamma_{ij}=\beta_i/T \delta_{ij}$ are diagonal. Moreover, the matrix $A_{ij}=0$ and the quadratic part of the effective potential vanishes, \ie, $U(R)=0$. From \cref{eq:Q0} we see that $\tilde q\itwo_{nm}$ is actually not needed, and we just have to solve \cref{eq:q1tilde} for
\beq
    \tilde q\ione_n = b_n^{-1} \left( \pd_R \bra n \ket \right) \left[  W' Q\izero + T \pd_R Q\izero \right] \, ,
\eeq
and plug this back into \cref{eq:Q0} together with $q\ione_n$ given in \cref{eq:wickthm}. This yields an effective FP equation for the tracer particle in the form of \cref{eq:effective_FP}, with the drift and diffusion coefficients given by
\beq 
    \mu(R) = \mu_0(R) + D'(R),\qquad \mu_0(R) = - \left[ 1-\chi T M(R)\right] W'(R), \qquad
    D(R) = T - \chi T  M(R) ,
    \label{eq:coefficients_linear_reactive}
\eeq 
where we introduced [see \cref{eq:u_and_v,eq_tn_def,eq_sn_def} for the definitions of $v_n$, $t_n$, and $s_n$, respectively]
\beq
    M(R) \equiv \frac{t_n}{b_n} \left( \pd_R \bra n \ket \right)  = \sum_n \frac{t_n^2}{\beta_n b_n} = T \sum_n \frac{\left[ \pd_R v_n(R)\right]^2}{\beta^2_n L_n} \, .
    \label{eq:M(R)}
\eeq
We note that the latter generalizes the quantity $m(R)$ defined in Ref.~\cite{gross_dynamics_2021} for the particular case in which $\Kcal=\mathbb{1}$, and the field Hamiltonian is of the Landau-Ginzburg type (see \cref{par:application_LG}).
The steady-state distribution of the resulting FPE is by construction the one given in \cref{eq:marginal_stationary} -- see the discussion after \cref{eq:final_moments}.
Finally, in the absence of quadratic couplings, the effective potential in \cref{eq:W(R)} reduces to
\begin{equation}
    W(R) = -\frac{1}{2} \sum_n \frac{v_n^2(R)}{\beta_n} - h_1 \sum_n \frac{v_n(R)}{\beta_n} [\sigma_n(0)+\sigma_n(L)] + \T{const.} \, .
    \label{eq:potential_pure_linear}
\end{equation}
It is useful to note that $v_n\sim\Ocal(h)$, hence $M(R)$ and $W(R)$ are of $\Ocal(h^2)$.

\subsubsection{Quadratic coupling}
\label{par:reactive_quadratic}
In this case the linear couplings (including boundary fields) are set equal to zero, $h=h_1=0$, so that also $s_n=t_n=0$ and the linear part of the effective potential vanishes, \ie, $W(R)=0$. In turn, this implies that the one-point cumulants $\bra n \ket$ in \cref{eq:wickthm} vanish as well. Starting from \cref{eq:q2tilde}, we denote the quantity on the r.h.s. as
\beq
    \Omega_{nm} \equiv \left( \pd_R \bra nm \ket \right) \left[ U' Q\izero + T \pd_R Q\izero \right] \, ,
    \label{eq:Omega}
\eeq
and thus rewrite
\begin{equation}
    B_{nj} \tilde q\itwo_{jm} + B_{mj} \tilde q\itwo_{jn} = \Omega_{nm} \, .
    \label{eq:quad_lyapunov}
\end{equation}
The latter is a Lyapunov matrix equation (see also \cref{app:lyapunov}), which, despite the fact that $B_{ij}$, $\Omega_{ij}$ and $\tilde{q}\itwo_{ij}$ are symmetric matrices, does not admit a straightforward analytic solution because in general $B\Omega \neq \Omega B$. Using the definition of $B_{ij}$ in \cref{eq:def_B}, we can however rephrase \cref{eq:quad_lyapunov} as (no sum over $n,m$ is intended)
\beq
    (b_n+b_m) \tilde q\itwo_{nm} = \Omega_{nm} - c_{nk} \tilde q\itwo_{km} - c_{mk} \tilde q\itwo_{nk} \, ,
    \label{eq:dyson_like}
\eeq
which can be taken as a starting point for a recursive solution in orders of $c$, noting that $c_{ij} \sim \cor{O}(c)$. Equation \eqref{eq:dyson_like} resembles a Dyson sum, but it does not admit a straightforward exact resummation; however, its ``bare'' solution
\beq
    \tilde q\itwo_{nm} \simeq \frac{\Omega_{nm}}{b_n+b_m} + \cor{O}(c^2)
    \label{eq:q2tilde_bare}
\eeq
can already be adopted as a small-$c$ approximation, which retains by construction the property of vanishing at equilibrium (\ie, in correspondence of the stationary distribution in \cref{eq:marginal_stationary}).
Note that the correction in \cref{eq:q2tilde_bare} is at least of $\cor{O}(c^2)$, because $\Omega_{nm}$ itself contains contributions at least of $\cor{O}(c)$ -- see \cref{eq:Omega}.
Note also that $c$ is a dimensionful quantity, so that the small-$c$ regime is in fact defined by the smallness of a suitable dimensionless counterpart of the coupling parameter. The particular expression of this parameter is model-dependent -- see, \eg, \cref{eq:c_dimensionless_LG,eq:c_dimensionless_GS} for the case of the Landau-Ginzburg and the Gompper-Schick models, respectively.

Inserting the approximate result reported in \cref{eq:q2tilde_bare} into the evolution equation \eqref{eq:Q0}, together with $q\itwo_{nm}$ found in \cref{eq:wickthm}, yields the effective FP equation
\begin{align} 
    \pd_t Q\izero =& \, \pd_R \left[ Q\izero \partial_R U + T \partial_R Q\izero \right] \label{eq:effective_FP_pure_quad} \\
    &-\chi T \partial_R \left[\frac{2}{1+V} \sum_{nm} \frac{A_{nm}^2 }{(b_n+b_m)\beta_n\beta_m} +\left(\frac{1}{1+V}\right)'  \sum_{nm} \frac{A_{nm} u_n u_m}{(b_n+b_m)\beta_n\beta_m}\right] \left[ Q\izero \partial_R U + T \partial_R Q\izero \right]+ \cor{O}(c^2) \n \, .
\end{align}
We can bring it to the form of \cref{eq:effective_FP} upon defining the drift and diffusion coefficients as 
\beq 
    \mu(R) = \mu_0(R) + D'(R),\qquad \mu_0(R) = - \left[ 1-\chi T M_2(R)\right] U'(R), \qquad
    D(R) = T - \chi T  M_2(R) ,
    \label{eq:coefficients_quadratic_reactive}
\eeq 
where we have introduced
\beq
    M_2(R) \equiv T \left[\frac{2}{1+V} \sum_{nm} \frac{A_{nm}^2 }{(b_n+b_m)\beta_n\beta_m} +\left(\frac{1}{1+V}\right)'  \sum_{nm} \frac{A_{nm} u_n u_m}{(b_n+b_m)\beta_n\beta_m}\right] \, .
    \label{eq:M2(R)}
\eeq
We recall that the expressions for $A_{nm}$ and $b_n$ are provided in \cref{eq:def_A,eq:b_and_c}, those for $\beta_n$ and $L_n$ can be found in \cref{par:examples}, while the potentials $U(R)$ and $V(R)$ are defined in \cref{eq:U_(R),eq:V(R)}.
At steady-state, the associated FPE is solved by $P_s$ given in \cref{eq:marginal_stationary} [again by construction, see the discussion after \cref{eq:final_moments}].

\subsubsection{Quadratic coupling with boundary fields}
\label{sec_quad_reac_bound}
In this case $h=0$ but $h_1 \neq 0$. The strategy is the same as in \cref{par:reactive_quadratic}, but now many new terms arise due to the boundary fields. The resulting effective FP equation takes the form of \cref{eq:effective_FP} upon defining
\beq 
    \mu(R) = \mu_0(R) + D'(R),\qquad \mu_0(R) = - \left[ 1-\chi T M_3(R)\right] \left[U'(R)+W'(R)\right], \qquad
    D(R) = T - \chi T  M_3(R) ,
    \label{eq:coefficients_quadratic_reactive_bf}
\eeq 
where
\begin{align}
    M_3(R) \equiv& M_2(R) +\frac{2T F_6}{1+V} \left[ \frac{G_4 \pd_R G_1 +G_1 \pd_R G_4/2}{1+V} + \frac{G_1 G_4 }{(1+V)'}\right]  +2 \left( G_6 F_4 \pd_R G_6 + G_6^2 F_5 -G_6 F_3 -F_2 \pd_R G_6 \right) \, ,
    \label{eq:M3(R)}
\end{align}
and where we introduced
\begin{align}
    &G_1(R) \equiv T \sum_i \frac{u_i \tau_i}{\beta_i} ,    && G_4(R) \equiv  \sum_i \frac{u_i^2}{b_i \beta_i},  &&& G_6(R) \equiv \frac{G_1(R)}{1+V(R)} , \n \\
    &F_2(R) \equiv T \sum_{nm} \frac{A_{nm} \tau_n u_m}{b_m\beta_n\beta_m},  && F_3(R) \equiv T \sum_{nm} \frac{A_{nm} \tau_n u_m'}{b_m\beta_n\beta_m},  &&& F_4(R) \equiv \sum_{nm} \frac{A_{nm} u_n u_m}{(b_n+b_m)\beta_n\beta_m} , \n \\
    &F_5(R) \equiv \sum_{nm} \frac{A_{nm}^2 }{(b_n+b_m)\beta_n\beta_m} ,   && F_6(R) \equiv \sum_{nm} \frac{A_{nm}  u_n b_m \tau_m  }{(b_n+b_m)\beta_n\beta_m} \, .
    \label{eq:F_G}
\end{align}
In particular, in the case in which $h=0$ but $h_1 \neq 0$, the effective potential $W(R)$ reduces to
\begin{equation}
    W(R) =  \frac{h_1^2}{2(1+V(R))} \left\lbrace \sum_{n} \frac{u_n(R) }{\beta_n} [\sigma_n(0)+\sigma_n(L)] \right\rbrace^2  + \T{const.} \, .
\end{equation}
Finally, the assumption $h=0$ can be released to address the most general case in which both a linear and a quadratic coupling are included. This gives rise to additional terms in the effective FP equation, which we report in \cref{app:mixed_case}.

\subsection{Passive case}
Here we consider the situation in which the parameter $\zeta=0$ in \cref{eq:langevin}. In this case the interaction between the field and the particle is non-reciprocal: the field influences the particle, but not vice-versa. As explained in \cref{sec_model}, this  models a specific kind of ``active'' particle driven by the stochastic process in \cref{eq:langevin_field}, which can be correlated and induce non-trivial dynamics and steady-states for the tracer. In this Section we disregard the possibility of adding boundary fields ($h_1=0$), which would be of limited physical significance; the relevant expressions for the evolution of the moments are then obtained by setting $s_n\equiv 0$ and $B_{nm} = b_n \delta_{nm}= \delta_{nm} L_n \beta_n/T_\phi$ in the coupled equations \eqref{eq:adiabatic}.

In the following we will consider the linear and the quadratic coupling cases separately. The comparison with previous results obtained in Refs.~\cite{dean_diffusion_2011,demery_perturbative_2011,demery_diffusion_2013} in the absence of confinement will be commented on in \cref{sec_discussion}, and further detailed in \cref{app:bulk}.

\subsubsection{Linear coupling}
In this case we also have $A_{ij}=0$. Grouping terms according to their order in $\chi$ in \cref{eq:Q1,eq:Q2} leads to
\begin{subequations}
\begin{align}
	q_{nm}\itwo &= \frac{T_\phi}{\beta_n}\delta_{nm} Q\izero \, ,  \\
    q_n\ione &= 0 \, , \label{eq:Q1_sol_linp}\\
    \tilde q\ione_n &=  -\frac{1}{b_n} \pd_R t_m q_{nm}\itwo = -\frac{T_\phi^2}{\beta^2_n L_n} \pd_R t_n Q\izero \, , \label{eq:Q1_tilde_sol_linp}
\end{align}\label{eq_Q_sol_linp}
\end{subequations}
where in the last result we used \cref{eq:Q0}.
Inserting \cref{eq:Q1_sol_linp,eq:Q1_tilde_sol_linp} back into \cref{eq:Q0} then renders the effective FPE for the reduced tracer distribution $Q\izero$. This FPE takes the same form as in \cref{eq:effective_FP}, but with drift and diffusion coefficients given by
\beq 
    \mu(R) = \onehalf D'(R),\qquad 
    D(R) = T_R + \chi T_\phi M(R) \, , 
    \label{eq:coefficients_linear_passive}
\eeq 
and with $M(R)$ defined as in \cref{eq:M(R)}, upon replacing $T$ with $T_\phi$. Explicit expressions of the function $M(R)$ for some selected models are provided in \cref{sec_applications} -- see \cref{eq:M(R)_LG_B,eq:M(R)_LG_A,eq:M(R)_GS}. 
Note that the correction to the diffusion coefficient $D(R)$ here has the opposite sign with respect to the one in the reactive case, see \cref{eq:coefficients_linear_reactive}.
One can easily check that the corresponding stationary distribution reads
\beq
    P_\T{s}(R) \propto D(R)^{-1/2} \, .
    \label{eq:stationary_dist_linear_passive}
\eeq
It is instructive to compare our results also to Refs.~\cite{dean_diffusion_2011,demery_perturbative_2011}, where a tracer particle linearly coupled to a fluctuating field has been analyzed in the \emph{bulk}. While a quantitative comparison requires specialization of the above results to periodic BCs and performing the bulk limit (see \cref{sec_eigenbasis} and the discussion in \cref{app:bulk}), we focus here on the sign of the correction term $M(R)$ to $D(R)$ [see \cref{eq:coefficients_linear_passive}]. 
In qualitative agreement with previous studies [see \cref{eq:M_lin_dean}], we find here that the effective diffusivity is generally reduced (enhanced) for a reactive (passive) tracer [see \cref{eq:coefficients_linear_reactive,eq:coefficients_linear_passive,eq:M(R)}]. Notably, it has been shown in Refs.\ \cite{dean_diffusion_2011,demery_perturbative_2011} that the diffusivity of a passive tracer coupled to a slowly relaxing field can even decrease below its bare value, an effect which is not captured within the adiabatic approximation.

\subsubsection{Quadratic coupling}
\label{par:passive_quadratic}
In this case $s_n=t_n=0$ in \cref{eq:adiabatic}, so that we only need the evolution equations for the even moments $Q^{(2n)}$. In particular, \cref{eq:Q0} implies that we need to determine the fourth moment $q\ifour_{ijnm}$ at the lowest order in $\chi$. However, in the passive case the system does not satisfy detailed balance, and thus the correct stationary distribution is not given by \cref{eq:stationary} but instead follows by solving the corresponding dynamical equation. Analogously to the approach leading to \cref{eq:adiabatic}, we can derive
\begin{align}
    \pd_t Q_{nmpq}\ifour =&  - \chi^{-1}\left[ B_{nl} Q_{mpql}\ifour + B_{ml} Q_{npql}\ifour + B_{pl} Q_{nmql}\ifour + B_{ql} Q_{nmpl}\ifour\right] \label{eq:Q4_dyn} \\
	&+ 2\chi^{-1}\left[ L_n\delta_{nm}Q_{pq}\itwo + L_m \delta_{mp}Q_{np}\itwo + L_p\delta_{pq}Q_{nm}\itwo + L_q\delta_{qm}Q_{np}\itwo + L_n\delta_{np}Q_{mp}\itwo + L_n\delta_{nq}Q_{mp}\itwo\right] +\Ocal(\chi^0)  \, . \n
\end{align}
Since  $B_{nm} = b_n \delta_{nm}$ is diagonal, from \cref{eq:Q2,eq:Q4_dyn} we find the $\cor{O}(\chi^0)$ solutions
\begin{align}
    q_{nm}\itwo &= \frac{T_\phi}{\beta_n}\delta_{nm} Q\izero, \label{eq:q2_pass_quad}\\
	q_{mnpq}\ifour &= 2T_\phi \frac{L_n \delta_{nm} q_{pq}\itwo + L_m\delta_{mp}q_{nq}\itwo + L_p\delta_{pq}q_{nm}\itwo + L_q\delta_{qm}q_{np}\itwo + L_n\delta_{np}q_{mq}\itwo + L_n\delta_{nq}q_{mp}\itwo}{L_n\beta_n + L_m\beta_m + L_p\beta_p + L_q\beta_q} \nonumber \\
		&= T_\phi^2 \left[ \frac{\delta_{nm}\delta_{pq}}{\beta_n \beta_q} + \frac{\delta_{mp}\delta_{nq} + \delta_{mq}\delta_{np}}{\beta_n\beta_m}\right] \, . \label{eq:q4_sol}
\end{align}
In order to evaluate \cref{eq:Q2} at $\Ocal(\chi^0)$, we need to determine $\pd_t q\itwo$ using \cref{eq:Q0,eq:q2_pass_quad}, which yields 
\begin{align}
    \tilde q_{nm}\itwo &= \frac{1}{c\,(b_n+b_m)} \left[-\frac{T_\phi^2}{\beta_n}\delta_{nm} \sum_p \pd_R \frac{A_{pp}}{\beta_p} Q\izero + \sum_{pq} \pd_R A_{pq} Q_{nmpq}\ifour\right] \nonumber \\
    &= \frac{T_\phi^2}{c\,(b_n+b_m)} \sum_p \pd_R \left[\frac{A_{pn}\delta_{mp}}{\beta_p\beta_n} + \frac{A_{pm}\delta_{np}}{\beta_p\beta_m}\right]Q\izero \, .
    \label{eq:q2tilde_pass_quad}
\end{align}
We recall that $A_{nm}$ and $b_n$ are reported in \cref{eq:def_A,eq:b_and_c}, while expressions for $\beta_n$ and $L_n$ for the various models can be found in \cref{par:examples}.
Finally, inserting \cref{eq:q2_pass_quad,eq:q2tilde_pass_quad} into \cref{eq:Q0} renders the effective FPE, valid to $\Ocal(\chi c^2)$ (see below),
\beq 
    \pd_t Q\izero =  \frac{T_\phi}{2} \pd_R V'(R) Q\izero + \pd_R^2 \left[T_R + 2\chi T_\phi^2  \sum_{nm} \frac{A_{nm}^2}{(b_n+b_m)\beta_n\beta_m}\right] Q\izero + \T{h.o.} \;.
    \label{eq:quadratic_passive}
\eeq
The potential $V(R)$ was defined in \cref{eq:V(R)}, and explicit expressions for some selected models are reported in \cref{sec_applications} below [see \cref{eq:V(R)_quad_Dirichlet,eq:V(R)_quad_Neumann,eq:V(R)_GS}]. We recognize 
\beq 
    \mu(R) = -\frac{T_\phi}{2} V'(R), \qquad 
    D(R) = T_R + \chi T_\phi M_4(R) \, ,
\label{eq:coefficients_quadratic_passive}
\eeq 
with
\begin{equation}
    M_4(R) = 2 T_\phi \sum_{nm} \frac{A_{nm}^2}{(b_n+b_m)\beta_n\beta_m} \, ,
    \label{eq:M4(R)}
\end{equation}
while the corresponding stationary distribution is formally given by 
\begin{equation}
    P_\T{s}(R) \propto \frac{1}{D(R)} \exp\left[\int_0^R \d r \frac{\mu(r)}{D(r)} \right] \equiv e^{- U_\T{eff}(R)/T_R} \, ,
    \label{eq:stationary_dist_generic}
\end{equation}
where we introduced the effective potential
\begin{equation}
    U_\T{eff}(R) \equiv -T_R \int_0^R \d r \frac{\mu(r)}{D(r)} +T_R \log D(R) \, .
    \label{eq:quadratic_passive_effective_potential}
\end{equation}

\subsubsection{Discussion}
\label{sec_discussion}
First of all, we note that the correction to the diffusion coefficient $D(R)$ in \cref{eq:coefficients_quadratic_passive} involves a (double) sum of positive terms, meaning that at $\Ocal(\chi)$ the diffusivity is enhanced due to the coupling with the (adiabatic) field. We remark that, in the bulk limit $L\to \infty$, we expect that $V(R)=\const$ by translational invariance: from \cref{eq:coefficients_quadratic_reactive,eq:M2(R)}, this implies that $D(R)$ is then \textit{reduced} in the reactive case. This qualitatively agrees with the findings in Ref.\ \cite{demery_diffusion_2013} (see \cref{eq:M_quad_dean}, while in the rest of \cref{app:bulk} we show that their agreement is also quantitative). Similarly to the linear case, one expects that, for a very slowly evolving field (non-adiabatic regime), the bulk diffusivity of a passive tracer can even fall below its bare value \cite{demery_diffusion_2013}.

Direct inspection of $V(R)$ for the case of the Landau-Ginzburg model with a simple quadratic coupling (see \cref{sec_LG_quad}) shows that actually $V(R)\to 0$ for $L\to \infty$, meaning that the correction to the diffusion coefficient in the reactive case [see \cref{eq:coefficients_quadratic_reactive,eq:M2(R)}] becomes equal in modulus (but opposite in sign) to that of the passive case, \cref{eq:coefficients_quadratic_passive}. The correct limiting procedure involves taking the limit $L\to\infty$ by keeping $\xi$ finite, which is why $V(R)$ presents instead a IR divergence for $L\to \infty$ in the critical models analyzed, \eg, in \cref{sec_GS} and in Ref.~\cite{gross_dynamics_2021}.
At criticality ($\xi=\infty$), the adiabatic approximation breaks down as $L\to\infty$ because the OP field becomes infinitely slow \cite{gross_dynamics_2021, Venturelli_2022}.

Secondly we note that, in contrast to the procedure of \cref{par:reactive_quadratic}, in the passive case there was no need to expand for small $c$ in order to obtain \cref{eq:quadratic_passive}; indeed, in the passive problem the field is agnostic to the value of $c$, since the latter merely sets the strength of the influence of the field on the dynamics of the particle. However, \cref{eq:quadratic_passive} does not admit a significant limit for large $c$, the drift term being $\propto c$ and the diffusion coefficient $\propto c^2$. This suggests that the \textit{effective} adiabaticity parameter in \cref{eq:quadratic_passive} is in fact proportional to $\chi c^2$, \ie, it involves both $\chi$ and the coupling constant $c$. Physically, by increasing $c$ in the Langevin equation \eqref{eq:langevin_part} for $R (t)$, one is actually speeding up the stochastic evolution of the tracer. In contrast, since the coupling constant $c$ does not enter the Langevin equation \eqref{eq:langevin_field} for the field $\phi$ in the passive case, the relaxation timescale of the latter remains the same. 
Accordingly, increasing $c$ eventually violates the assumption underlying the adiabatic approximation, namely that the field relaxes faster than the tracer particle; we must thus require $c$ to remain sufficiently small within the adiabatic approach.

A third remark is that, in the adiabatic regime, \cref{eq:quadratic_passive} retains the form of a Fokker-Planck equation even though the stochastic process $\propto \phi^2(R)$ which drives $R (t)$ is non-Gaussian.
Intuitively, this can be understood by considering the motion of the tracer along discrete time steps of length $\Delta t$ (which is assumed to be smaller than the tracer relaxation time). Since the field relaxation time can be made arbitrarily small in the adiabatic limit (e.g., of $\Ocal(\Delta t)$), the tracer will pick up a set of uncorrelated random noises $\propto \phi^2$ during its motion over several time steps. According to the central limit theorem, the sum of these noises assumes a Gaussian character.
Thus, this model is such that non-Markovian effects only appear at higher orders in the adiabaticity parameter $\chi$ (in contrast to other -- even simpler -- models featuring the square of a Gaussian process \cite{luczka_non-markovian_1995, Luczka_1988}). 

Next, in the case in which the coupling operator $\cor{K}_2$ in \cref{eq_Hamilt} is chosen to be the identity, the effective FP equation \eqref{eq:quadratic_passive} can be compared to Eq. (4.31) in Ref.~\cite{gross_dynamics_2021}. The latter was obtained within a small-$c$ expansion and following the standard Gardiner/Stratonovich adiabatic elimination method \cite{gardiner_adiabatic_1984,gardiner_stochastic_2009}. These methods generally assume (as we did in \cref{par:reactive}) that the fast variable relaxes to its equilibrium configuration around the slow variable, and then construct a perturbation series for small $\chi$ around this reference state. Note that no assumption in this sense has been invoked along the derivation of \cref{eq:quadratic_passive} outlined above. It turns out that the drift coefficient $\mu (R)$ in Eq.~(4.32a) of Ref.~\cite{gross_dynamics_2021} differs from \cref{eq:coefficients_quadratic_passive} by a spurious drift term (compatible with a Stratonovich interpretation of the noise). However, including this term would render a steady-state particle distribution $P_\T{s}(R)$ which coincides, up to $\cor{O}(c)$, with the stationary distribution in the presence of detailed balance given in \cref{eq:marginal_stationary} [see Eq. (4.33) in Ref.~\cite{gross_dynamics_2021}] -- this is not the correct stationary distribution, which is given instead by \cref{eq:stationary_dist_generic}. 

Following Ref.~\cite{gross_dynamics_2021} we eventually note that, upon defining from \cref{eqq_langevin_modes_R} an effective field-induced noise (see \cref{par:CLT})
\begin{equation}
    \Pi_c(R,t) \equiv - \sum_{nm} A_{nm}(R) \phi_n(t)\phi_m(t) \, ,
\end{equation}
the effective diffusion coefficient $D(R)$ reported in \cref{eq:coefficients_quadratic_passive} can be expressed as a Green-Kubo relation:
\beq 
    D(R) = T_R+ \int_{-\infty}^\infty \d t \bra \Pi_c(R,t)\Pi_c(R,0) \ket \, ,
    \label{eq:Kubo}
\eeq
where the average is intended over the stochastic noises in \cref{eq:langevin}.
Moreover, by using the definition of $A_{nm}$ in \cref{eq:def_A}, we can rewrite
\begin{equation}
    \Pi_c(z,t) \equiv -\frac{c}{2}\pd_z [\cor{K}_2\phi(z,t)]^2 \, ,
\end{equation}
which simplifies to $\Pi_c(z,t) \equiv -\frac{c}{2}\pd_z \phi^2(z,t)$ in the case where $\cor{K}_2=\mathbb{1}$ \cite{gross_dynamics_2021}.

\section{Application to specific models}
\label{sec_applications}
Here we apply the adiabatic elimination method developed in the previous Sections to the Landau-Ginzburg (LG) and the Gompper-Schick (GS) models, which have been introduced in \cref{par:examples}. In each of the cases considered below, we will discuss the stationary distribution $P_\T{s}(R)$ and the reduced diffusion coefficient
\begin{equation}
    D_r(R) \equiv \frac{D(R)-D_0}{\chi D_0} \, ,
    \label{eq:reduced_diff}
\end{equation}
where $D_0=T_R$ denotes the bare diffusion coefficient [which is of $\cor{O}(\chi^0)$]. Where possible, we provide analytic expressions for the reduced diffusivity or, correspondingly, for the functions $M_{1-4}(R)$, which are directly related to $D_r(R)$ via \cref{eq:coefficients_linear_reactive,eq:coefficients_quadratic_reactive,eq:coefficients_linear_passive,eq:coefficients_quadratic_passive}.
Henceforth, we choose units in which $T_R=1$ and focus on the case $T_\phi=T_R$.

\subsection{Landau-Ginzburg model}
\label{par:application_LG}
Turning first to the LG Hamiltonian, we consider in the following both dissipative (model A) and conserved (model B) OP dynamics in the presence of simple linear and quadratic couplings (\ie, $\cor{K}_1=\cor{K}_2 = \mathbb{1}$). Since many results for this model at the critical point (\ie, $\tau=0$) have been reported in Ref.~\cite{gross_dynamics_2021}, we will focus here on the case $\tau\neq 0$, for which the field acquires a finite correlation length $\xi=1/\sqrt{\tau}$.

\begin{figure}
    \centering
    \subfigure[]{
        \includegraphics[width=0.32\textwidth]{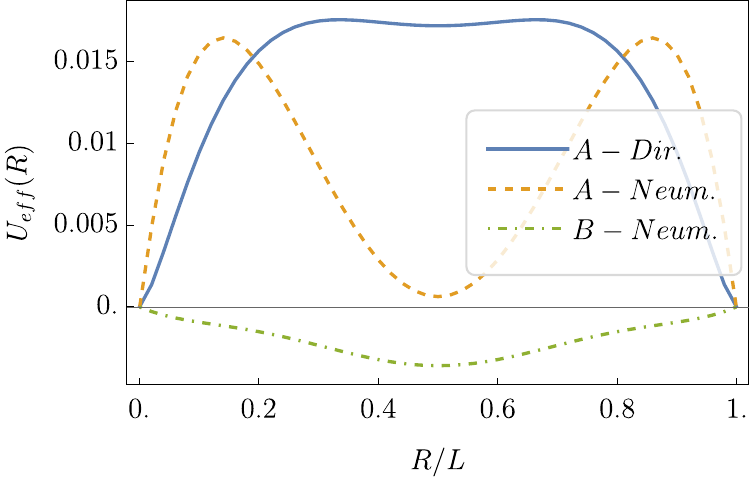}
        \label{fig:LG_quad_pass_pot}
    }
    \subfigure[]{
        \includegraphics[width=0.32\textwidth]{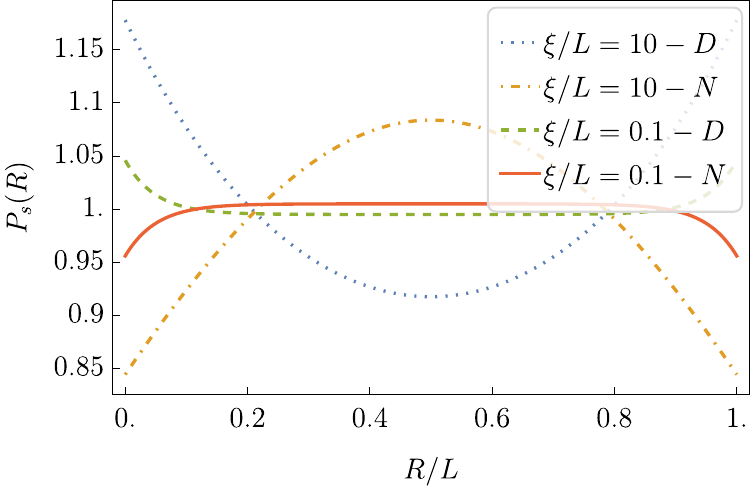}
    }
    \subfigure[]{
        \includegraphics[width=0.32\textwidth]{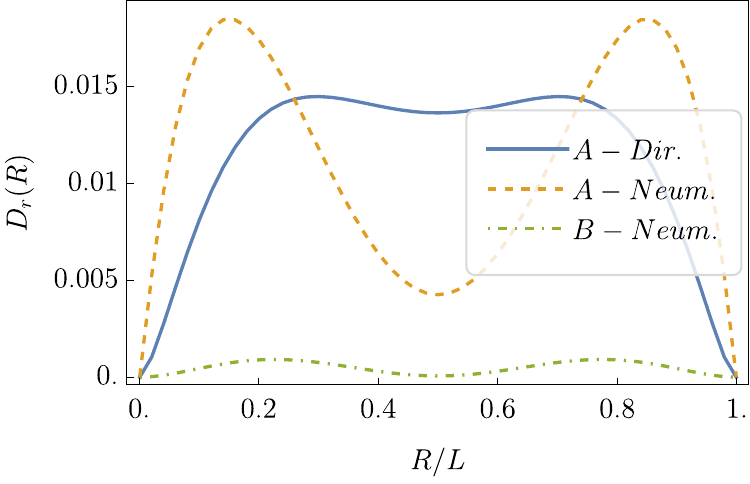}
    }
    \subfigure[]{
        \includegraphics[width=0.32\textwidth]{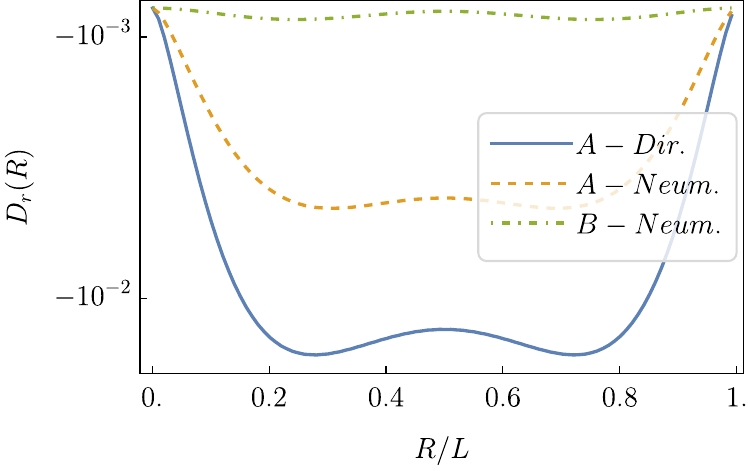}
    }
    \caption{Landau-Ginzburg model with quadratic coupling. (a) Effective potential $U_\T{eff}(R)$ given in \cref{eq:quadratic_passive_effective_potential} [with $V(R)$ given in \cref{eq:V(R)_quad_Dirichlet,eq:V(R)_quad_Neumann}] for a passive tracer with model A/B dynamics and Dirichlet/Neumann \bcs. The correlation length is set to $\xi/L=1$ and $\tilde\chi=1$. (b) Normalized stationary distribution $P_s(R)$ for a reactive tracer given in \cref{eq:marginal_stationary} and two distinct values of the correlation length $\xi$. The distribution becomes more flat upon decreasing $\xi$, with some residual structure in the proximity of the boundaries (limited in a layer of width $\xi$). (c,d) Reduced diffusion coefficient $D_r(R)$ [see \cref{eq:reduced_diff}], with $\xi/L=1$, for the (c) passive and (d) reactive case [see \cref{eq:coefficients_quadratic_passive,eq:coefficients_quadratic_reactive}]. In the plots we set $h_1=0$, while we used $\varkappa_c=0.1$ for the passive case and 
    $\varkappa_c=1$ for the reactive case.
    }
    \label{fig:LG_quad}
\end{figure}

\subsubsection{Quadratic coupling}
\label{sec_LG_quad}
The choice of a simple (non-derivative) coupling $\propto \phi^2(R)$ gives $u_n = \sqrt{c} \sigma_n$ [see \cref{eq:u_and_v}]. The potential $V(R)$ introduced in \cref{eq:V(R)} characterizes most static and dynamical properties in the quadratic case [see \cref{par:reactive_quadratic,par:passive_quadratic}]; here it becomes 
\beq
    V(R) = c \sum_n \frac{\sigma_n^2(R)}{\beta_n} \, ,
    \label{eq:V(R)_LG_pure}
\eeq
where one can recognize $V(R) = \frac{c}{T}C_\phi(R,R)$ in terms of the OP correlator [see Eq. (3.20) in \cite{gross_dynamics_2021}]. This quantity can in fact be computed even off-criticality, as detailed in \cref{app:details_series}: for Dirichlet \bcs, we find
\begin{equation}
    V(R) = c \xi \,  \T{csch}(L/\xi) \sinh(R/\xi) \sinh\left(\frac{L-R}{\xi}\right)  \, .
    \label{eq:V(R)_quad_Dirichlet}
\end{equation}
For $\xi \to \infty$, the above expression reduces to a quadratic function in $R$ \cite{gross_dynamics_2021}, while it decays to zero upon decreasing $\xi$.
Choosing Neumann BCs (including the zero mode), we find instead
\begin{equation}
    V(R) =\frac{c \xi}{2}\, \T{csch}(L/\xi) \left[ \cosh (L
    /\xi)  + \cosh \left(\frac{L-2R}{\xi}\right) \right] \, ,
    \label{eq:V(R)_quad_Neumann}
\end{equation}
whose limit for $\xi \to \infty$ gives a $R$-independent diverging constant (which is removed by excluding the zero mode). Both \cref{eq:V(R)_quad_Dirichlet,eq:V(R)_quad_Neumann} show that $V(R)\to 0$ in the bulk limit $L\to \infty$ (with $\xi$ kept finite -- see the discussion in \cref{sec_discussion}).

Returning to the issue of finding a dimensionless counterpart of $c$ (see \cref{par:reactive_quadratic}), we note that dimensional analysis renders $[c]=[L]^{-1}$ in units of length. 
Following Ref.\ \cite{gross_dynamics_2021} (see Eq. (4.35) therein), we define the corresponding dimensionless parameter
\begin{equation}
    \varkappa_c \equiv c L \, ,
    \label{eq:c_dimensionless_LG}
\end{equation} 
remarking that, sufficiently far from criticality, replacing $L$ by $\xi$ would render an equally admissible choice.

Finally, we can compute the stationary potential $W(R)$ in \cref{eq:W(R)}. In the quadratic case we have $h=0$, so a non-vanishing $W(R)$ is only obtained in the presence of boundary fields ($h_1\neq 0$). This case is described by capillary \bcs (see \cref{eq_bcs}), for which
\begin{equation}
    W(R) = \frac{c}{2(1+V(R))} \left\lbrace \frac{h_1  \xi}{\sinh(L/\xi)}\left[ \cosh \left(\frac{L-R}{\xi}\right)+\cosh \left(\frac{R}{\xi}\right)  \right] \right\rbrace^2 \, .
    \label{eq:W(R)_quad}
\end{equation}
Again, this form suggests to introduce a dimensionless coupling describing the significance of boundary fields as 
\begin{equation}
    \varkappa_{h_1}\equiv h_1 \sqrt{L}.
    \label{eq:dimensionless_bf}
\end{equation}
For a quadratically coupled tracer, it is in general not possible to obtain analytical expressions for the drift and diffusion coefficients reported in \cref{eq:coefficients_quadratic_reactive,eq:coefficients_quadratic_passive}, apart from $\mu(R)$ in the passive case.
In particular, the stationary distribution in the passive case [which is given in \cref{eq:stationary_dist_generic} in terms of $D(R)$] has to be evaluated numerically, while $P_s(R)$ for the reactive case is available in explicit form via \cref{eq:marginal_stationary}. Both distributions are plotted in the first two panels of \cref{fig:LG_quad}. 

In the reactive case, the stationary distribution is independent of the type of dynamics (model A or B) due to detailed balance. 
Furthermore, the probability density of the tracer particle is peaked at the boundaries for Dirichlet \bcs, or at the center of the interval for Neumann \bcs, but both these features become less pronounced as we move away from the critical point, \ie, upon decreasing $\xi$. The difference between model A and B, however, becomes more evident in the passive case: in particular, for model A and Neumann \bcs, the stationary distribution is no longer unimodal (see \cref{fig:LG_quad_pass_pot}). 
Similar features are shared by the (reduced) diffusion coefficients $D_r(R)$ in the passive and reactive case, which are plotted in the bottom row of \cref{fig:LG_quad}.
Notably, the effective diffusivity $D_r$ is enhanced in the passive case due to the additional noise provided by the OP field.
By contrast, in the reactive case, the diffusivity is reduced, which can be understood as a consequence of the suppression of OP fluctuations due to the quadratic coupling. 
These findings are in qualitative agreement with the behavior of a tracer in a bulk medium \cite{demery_diffusion_2013}.

\begin{figure}
    \centering
    \subfigure[]{
        \includegraphics[width=0.31\textwidth]{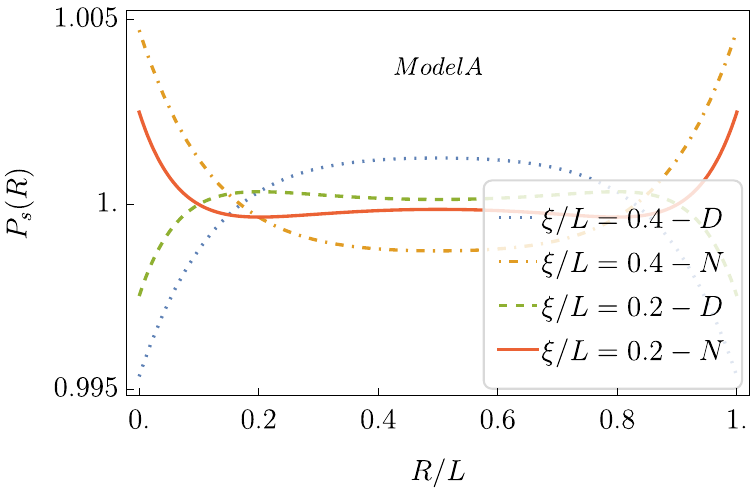}
    }
    \subfigure[]{
        \includegraphics[width=0.31\textwidth]{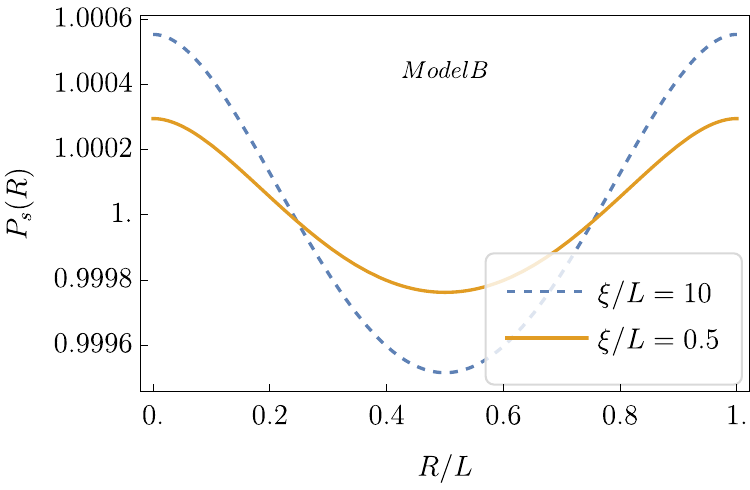}
    }
    \subfigure[]{
        \includegraphics[width=0.31\textwidth]{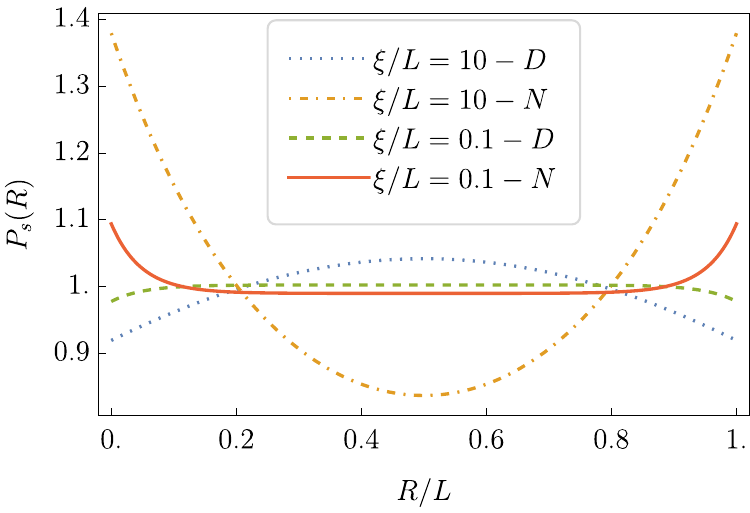}
    }
    \subfigure[]{
        \includegraphics[width=0.31\textwidth]{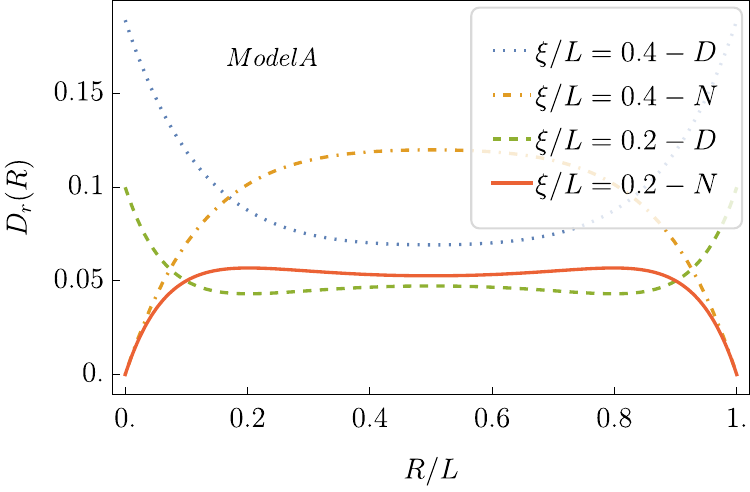}
    }
    \subfigure[]{
        \includegraphics[width=0.31\textwidth]{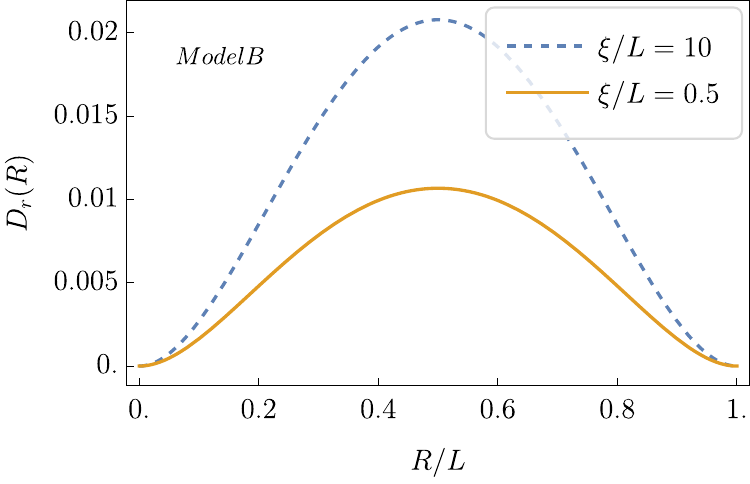}
    }
    \caption{Landau-Ginzburg model with linear coupling. Stationary distribution $P_s(R)$ given in \cref{eq:stationary_dist_linear_passive} [with the function $M(R)$ given in \cref{eq:M(R)_LG_B,eq:M(R)_LG_A}] for a passive tracer with (a) model A and (b) model B field dynamics (Neumann \bcs only), using $\tilde\chi=0.1$. (c) Stationary distribution $P_s(R)$ given in \cref{eq:marginal_stationary} [with the function $W(R)$ given in \cref{eq:W(R)_lin}] for a reactive tracer and two different correlation lengths. (d)-(e) Reduced diffusion coefficient $D_r(R)$ [see \cref{eq:reduced_diff}] for a passive tracer with model A/B dynamics (which is equal in magnitude and opposite to that of the reactive case). In the plots we used $\varkappa_h=1$ and $h_1=0$.}
    \label{fig:LG_lin}
\end{figure}

\subsubsection{Linear coupling}
\label{par:LG_linear_application}
The analysis of the LG model with a linearly coupled particle was presented in Ref.~\cite{gross_dynamics_2021} in the case of a critical field ($\tau=0$). Here we extend it to the off-critical case, where the field acquires a finite correlation length $\xi=1/\sqrt{\tau}$.

We start by computing the quantity $M(R)$ in \cref{eq:M(R)}, which characterizes the drift and diffusion coefficients both in the reactive and passive cases [see \cref{eq:coefficients_linear_reactive,eq:coefficients_linear_passive}, respectively], and, in particular, it determines the stationary distribution in the passive case [see \cref{eq:coefficients_linear_passive,eq:stationary_dist_linear_passive}]. Noting that the choice of a simple (non-derivative) coupling $\propto \phi(R)$ gives $v_n = h \sigma_n$ [see \cref{eq:u_and_v}], we find in the case of model B with Neumann BCs (see \cref{app:details_series})
\begin{align}
    M(R)_{\T{Neum}} = -2 \left[\frac{h\xi }{2\sinh (L/\xi)}\right]^2 \Bigg\lbrace L \sinh^2 \frac{R}{\xi}+\sinh \frac{L}{\xi} \cdot \Bigg[ R \sinh{\frac{L-2R}{\xi}} -\xi \sinh{\frac{L-R}{\xi}}\sinh{\frac{R}{\xi}} \Bigg] \Bigg\rbrace \label{eq:M(R)_LG_B}
\end{align}
(recall that model B dynamics with a globally conserved OP field is incompatible with Dirichlet BCs -- see \cref{par:examples}).
Similar expressions hold for model A (see \cref{app:details_series}):
\begin{align}
    M(R)_{\T{Dir}} =& \, \frac{h^2}{8 \sinh^2(L/\xi)}\Bigg\lbrace 2(R-L) \cosh \left(\frac{2R}{\xi}\right) -2R \cosh \left(\frac{2(L-R)}{\xi}\right) -2L \label{eq:M(R)_LG_A} \\
    &+\xi \left[  \sinh \left(\frac{2R}{\xi}\right) + \sinh \left(\frac{2L}{\xi}\right) + \sinh \left(\frac{2(L-R)}{\xi}\right) \right] \Bigg\rbrace  , \n \\
    M(R)_{\T{Neum}} =& \, \frac{h^2}{8 \sinh^2(L/\xi)} \Bigg\lbrace \xi \sinh \left(\frac{2L}{\xi}\right) - 2\sinh \left(\frac{L}{\xi}\right) \left[ \xi \cosh \left(\frac{L-2R}{\xi}\right) -2R \sinh \left(\frac{L-2R}{\xi}\right)  \right] +4L \sinh^2 \left(\frac{R}{\xi}\right) \Bigg\rbrace  .\n
\end{align}
The critical point ($\tau=0$) has already been considered in Ref.~\cite{gross_dynamics_2021}, in which case the function $M(R)$ simplifies to a polynomial form $m(R)$ (see, e.g., Eq.~(3.26) in Ref.\ \cite{gross_dynamics_2021}).
Note, however, that it is not directly possible to recover $m(R)$ from $M(R)$ by simply taking the limit for $\xi\to\infty$ in the latter: inspection of the relevant series for Neumann \bcs reveals the presence of a constant, diverging zero mode which should be manually removed in order to yield meaningful results (see \cref{app:details_series}). For any finite value of $\xi$, instead, the behavior of this zero mode is regular.

The stationary potential $W(R)$ given in \cref{eq:potential_pure_linear} reduces to
\begin{align}
    W(R) = -\frac{h^2 }{2c} V(R) -\frac{h h_1  \xi}{\sinh(L/\xi)}\left[ \cosh \left(\frac{L-R}{\xi}\right)+\cosh \left(\frac{R}{\xi}\right)  \right] \, .
    \label{eq:W(R)_lin}
\end{align}
Here the function $V(R)$ formally coincides with that given in \cref{eq:V(R)_quad_Dirichlet,eq:V(R)_quad_Neumann} (for the case of Dirichlet/Neumann \bcs, respectively). Since $V(R)$ is proportional to $c$, the constant $c$ (which is zero in the linear case) does not enter \cref{eq:W(R)_lin}. The term proportional to $h_1$, which contains the effect of the boundary fields, implies the use of Neumann modes [\cref{eq_eigenf_Nbc}] for the OP. In this case, we can identify the dimensionless coupling $\varkappa_{h_1}$ to the boundary fields as in \cref{eq:dimensionless_bf}, and the dimensionless linear coupling to the tracer particle as
\begin{equation}
    \varkappa_h = h \sqrt{L} \, .
\end{equation}

The stationary distributions for the passive [\cref{eq:stationary_dist_linear_passive}] and reactive [\cref{eq:marginal_stationary}] cases are plotted in \cref{fig:LG_lin} [panels (a-c)]. The case of passive model A is particularly interesting, as the qualitative character of the stationary distribution (\ie, the fact that it is peaked either at the boundaries or in the middle of the interval) drastically changes as we approach the critical point. This is qualitatively confirmed by numerical simulations performed along the lines of Ref.~\cite{gross_dynamics_2021} (not shown), and it is reflected in the behavior of the (reduced) diffusion coefficient $D_r(R)$ which is plotted in panels (d-e) of \cref{fig:LG_lin}.
We find that the effective diffusivity $D_r$ is enhanced for a passive tracer when the OP is near-critical ($\xi\gg L$), which is due to the additional noise provided by the field. Notably, before $D_r$ vanishes as $\xi\to 0$,
it acquires a bimodal character: indeed, the effect of the boundary conditions has a spatial extension of $\mathcal{O}(\xi)$, which does not reach the center of the interval ($R=L/2$) when $\xi$ is sufficiently small.
In the reactive case, we obtain a $D_r$ equal in magnitude but of opposite sign to the one in the passive case (not shown). The reduction of the diffusivity stems from to the creation of a OP ``halo'' around the tracer as a direct consequence of the reactive coupling. The qualitative trends observed here are in agreement with the findings in Refs.\ \cite{dean_diffusion_2011, demery_perturbative_2011}.

\begin{figure}
    \centering
    \subfigure[]{
        \includegraphics[width=0.32\textwidth]{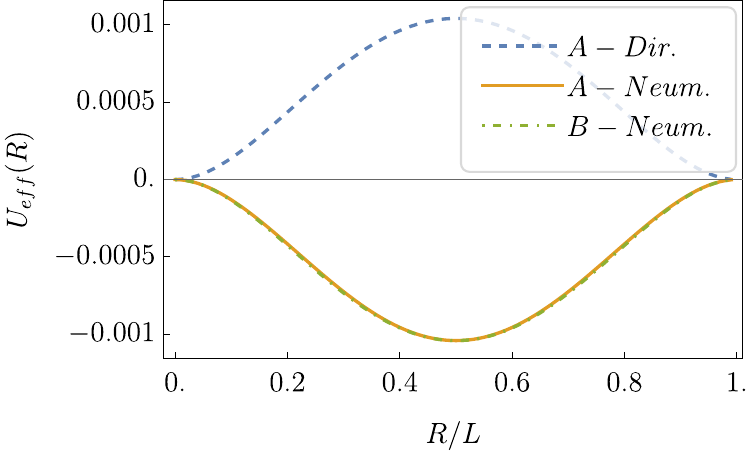}
    }
    \subfigure[]{
        \includegraphics[width=0.32\textwidth]{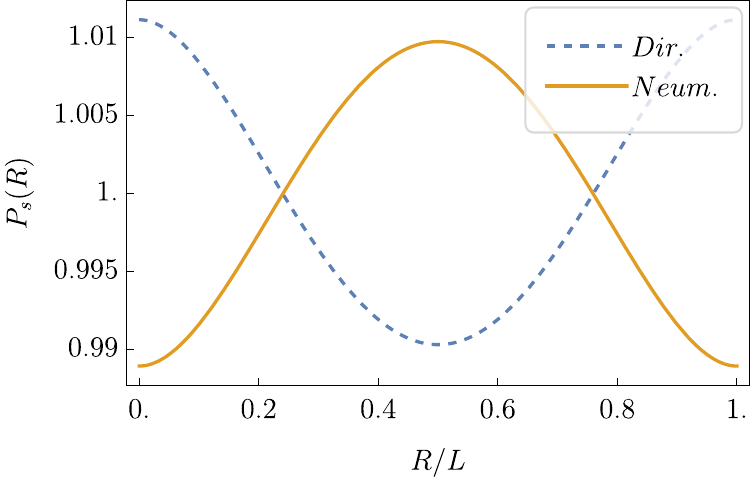}
    }
    \subfigure[]{
        \includegraphics[width=0.32\textwidth]{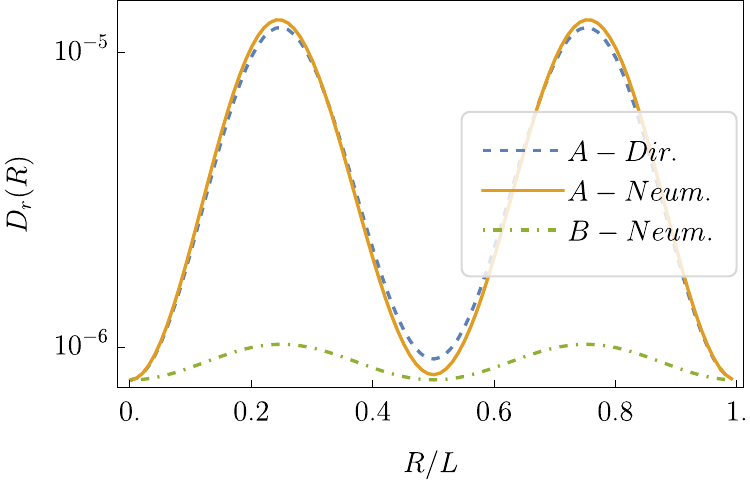}
    }
    \subfigure[]{
        \includegraphics[width=0.32\textwidth]{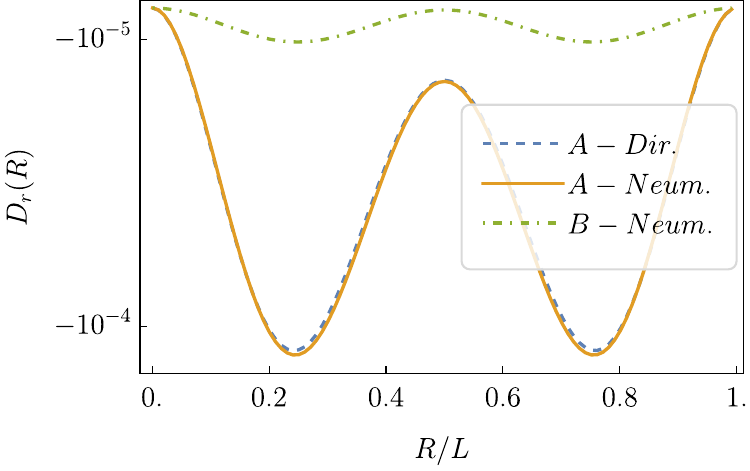}
    }
    \caption{Critical Gompper-Schick model with quadratic coupling. (a) Effective potential $U_\T{eff}(R)$ given in \cref{eq:quadratic_passive_effective_potential} [with $V(R)$ given in \cref{eq:V(R)_GS}] for a passive tracer with model A/B dynamics and Dirichlet and Neumann \bcs ($\chit=1$). (b) Normalized stationary distribution $P_s(R)$ for a reactive tracer given in \cref{eq:marginal_stationary}. (c,d) Reduced diffusion coefficient $D_r(R)$ [see \cref{eq:reduced_diff}] for the (c) passive and (d) reactive case. In the plots we set $h_1=0$, and used $\varkappa_c\ut{(GS)}=0.1$ in the passive and $\varkappa_c\ut{(GS)}=1$ in the reactive case [see \cref{eq:c_dimensionless_GS}].}
    \label{fig:GS_quad}
\end{figure}
\begin{figure}
    \centering
    \subfigure[]{
        \includegraphics[width=0.31\textwidth]{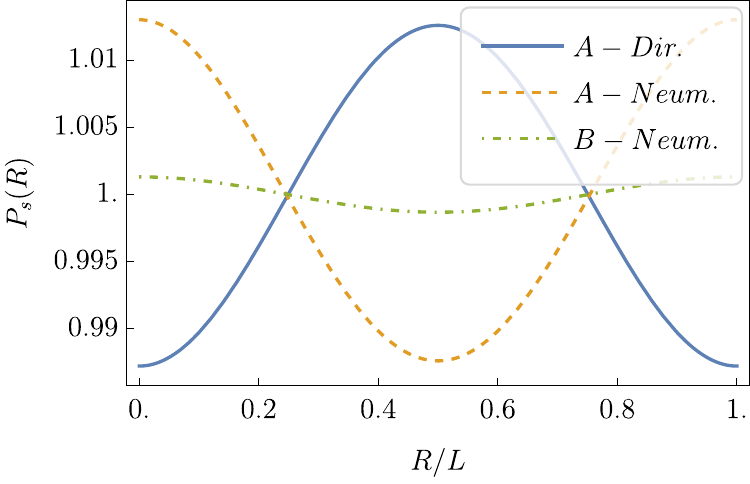}
    }
    \subfigure[]{
        \includegraphics[width=0.31\textwidth]{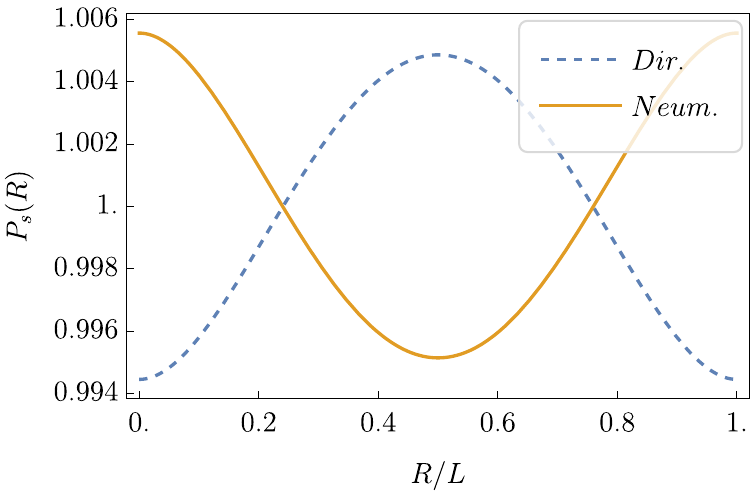}
    }
    \subfigure[]{
        \includegraphics[width=0.31\textwidth]{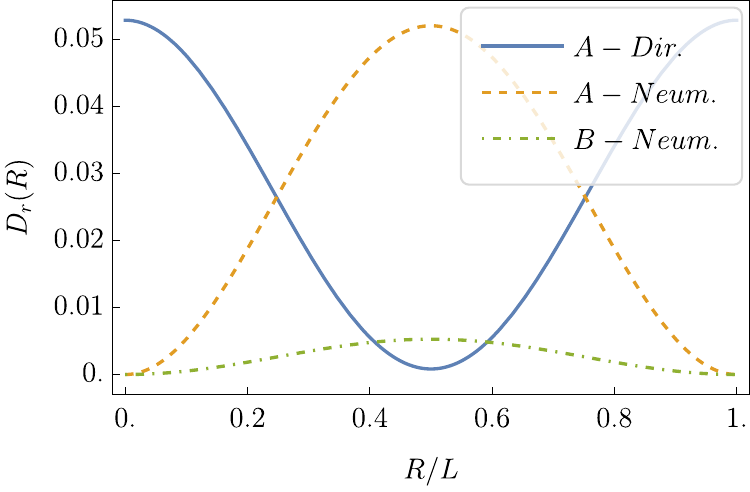}
    }
    \caption{Critical Gompper-Schick model with linear coupling. (a) Stationary distribution $P_s(R)$ given in \cref{eq:stationary_dist_linear_passive} [with the function $M(R)$ given in \cref{eq:M(R)_GS}] for a passive tracer with model A/B dynamics ($\tilde \chi =1$). (b) Stationary distribution $P_s(R)$ for a reactive tracer given in \cref{eq:marginal_stationary} [with the function $W(R)$ given in \cref{eq:W(R)_GS_lin}]. (c) Reduced diffusion coefficient $D_r(R)$ [see \cref{eq:reduced_diff}] for a passive tracer with model A/B dynamics. Note that $D_r$ is equal in magnitude and opposite to the one of the reactive case (not shown). In the plots we set $h_1=0$, while we used $\varkappa_h^\T{(GS)}= 5$ in the passive case and $\varkappa_h^\T{(GS)}= 1$ in the reactive case [see \cref{eq:h_dimensionless_GS}].}
    \label{fig:GS_lin}
\end{figure}

\subsection{Critical Gompper-Schick model} 
\label{sec_GS}
Now we turn to the critical Gompper-Schick model, which differs from the LG model by the presence of higher order derivatives in its field Hamiltonian [see \cref{eq:free_energy_GS}]. We focus on the critical point, which corresponds to $\tau=0$. 
In order to make the analytical computation of the stationary potentials appearing in the effective FPEs tractable, we will retain here only $\Delta(\rv) = c_4 \nabla_\rv^4$ in \cref{eq:operators_GS}, and consequently $\beta_n = c_4 k_n^4$. This is expected to already capture the essential physics and the main differences with respect to the LG Hamiltonian. We will again consider simple linear and quadratic couplings, \ie, $\cor{K}_1=\cor{K}_2 = \mathbb{1}$. We recall that, in the reactive case, the steady-state distribution is given by \cref{eq:marginal_stationary}. The stationary distribution in the passive linear case is reported in \cref{eq:stationary_dist_linear_passive} (and thus it is possible to write it analytically in terms of $M(R)$ computed below), while that of the passive quadratic case [which is formally stated in \cref{eq:stationary_dist_generic}] does not admit a straightforward analytic expression.

For the GS Hamiltonian, the effective potential defined in \cref{eq:V(R)} for the quadratic coupling is given by (see \cref{app:details_series})
\begin{equation}
    V(R) = \frac{c}{L c_4} \left(\frac{L}{\pi}\right)^4 \left[ f_4(0) \mp f_4\left( \frac{2\pi R}{L} \right) \right] \, ,
    \label{eq:V(R)_GS}
\end{equation}
where the function $f_n(x)$ is a $n$-th order polynomial (see \cref{eq:f_n_def}), and the signs $\mp$ correspond to Dirichlet or Neumann \bcs, respectively (we have subtracted the zero mode in the Neumann case). Note that $V(R)$ in \cref{eq:V(R)_GS} does not admit a finite limit for $L\to \infty$ due to IR divergences related to criticality (see also the discussion in \cref{sec_discussion}).

In this case, \cref{eq:V(R)_GS} suggests to identify a dimensionless coupling constant as
\beq
    \varkappa_c^\T{(GS)} \equiv \frac{cL^3}{c_4}   \, ,
    \label{eq:c_dimensionless_GS}
\eeq
where we note the length dimension $[c_4]=[L]^{2}$.
Similarly, in the purely linear case ($c=0$) we obtain
\begin{align}
    W(R) =& -\frac{h^2}{2L c_4} \left(\frac{L}{\pi}\right)^4 \left[ f_4(0) \mp f_4\left( \frac{2\pi R}{L} \right) \right] \n\\
    & -\frac{2h h_1}{L c_4} \left(\frac{L}{\pi}\right)^4 \left[ f_4\left( \frac{\pi R}{L} \right) + \frac12  f_4\left( \pi(1+R/L) \right) + \frac12  f_4\left( \pi(1-R/L) \right) \right] \, , \label{eq:W(R)_GS_lin}
\end{align}
while in the quadratic case with boundary fields ($h=0$ and $h_1\neq 0$) we find instead
\begin{equation}
    W(R) = \frac{h_1^2 c}{2(1+V(R))} \left\lbrace\frac{2}{L c_4} \left(\frac{L}{\pi}\right)^4 \left[ f_4\left( \frac{\pi R}{L} \right) + \frac12  f_4\left( \pi(1+R/L) \right) + \frac12  f_4\left( \pi(1-R/L) \right) \right]\right\rbrace^2 \, . \label{eq:W(R)_GS_quad}
\end{equation}
In the last two expressions, the part proportional to $h_1$, which encodes the effect of boundary fields, is only present for Neumann \bcs. Here, \cref{eq:W(R)_GS_lin} suggests to define the dimensionless coupling constant
\beq
    \varkappa_h^\T{(GS)} \equiv \frac{h L^{3/2}}{\sqrt{c_4}}  \, .
\label{eq:h_dimensionless_GS}\eeq
Analogously, a dimensionless coupling quantifying the importance of the contribution of the boundary fields $h_1$ can be introduced as
\beq
    \varkappa_{h_1}^\T{(GS)} \equiv \frac{h_1 L^{3/2}}{\sqrt{c_4}}  \, ,
\eeq
such that $W(R)$ in \cref{eq:W(R)_GS_quad} is $\propto \varkappa_c \, (\varkappa_{h_1}\ut{(GS)})^2$.
Finally, the dynamical coefficient $M(R)$ of the linear case reads
\begin{equation}
    M(R) = \frac{h^2}{L c_4^2} \left(\frac{L}{\pi}\right)^{6+2a} \left[ f_{6+2a}(0) \pm f_{6+2a}\left( \frac{2\pi R}{L} \right) \right] \, ,
    \label{eq:M(R)_GS}
\end{equation}
where $a=0,1$ marks the distinction between model A/B dynamics. 

The stationary distributions and diffusion coefficients corresponding to the quadratic or linear coupling cases are reported in \cref{fig:GS_quad,fig:GS_lin}, respectively.
The (reduced) diffusion coefficient $D_r(R)$ of a quadratically coupled tracer is markedly bimodal (see \cref{fig:GS_quad}), as the contribution of the field to the diffusion of the particle nearly vanishes both near the boundaries and at the center of the interval.  
Similarly to the LG model, both for a quadratically and a linearly coupled tracer (see \cref{fig:LG_quad,fig:LG_lin}), the diffusivity is enhanced (reduced) in the case of a passive (reactive) tracer. 
As before, one may understand these trends as a consequence of the additional field-induced noise imposed on a passive tracer, and of the slowing effect of the OP halo created around a reactive tracer.

\section{Summary and conclusions}
\label{sec_conclusions}
In this work we presented a simple and systematic procedure to study the effective dynamics of a tracer particle coupled to a confined correlated medium within the adiabatic approximation. The medium is modeled as a scalar order parameter $\phi(\rv,t)$ evolving under dissipative or conserved Langevin dynamics within the Gaussian approximation, and it is confined by the imposition of suitable boundary conditions at the ends of a one-dimensional interval. The particle at position $\Rv (t)$ undergoes a stochastic motion and it is subject to reflective \bcs. The interaction between the field and the particle is modeled by the addition of linear or quadratic coupling terms in the Hamiltonian in \cref{eq_Hamilt}: in the former case the field $\phi(\Rv,t)$ (or its derivatives) are enhanced in the vicinity of the tracer particle, while in the latter the correlations of the field are suppressed. 
If detailed balance is satisfied by the dynamics [\ie, if $\zeta=1$ and $T_R=T_\phi$ in \cref{eq:langevin,eq:noise_field}], we call the particle \textit{reactive}, as its back-reaction on the medium is taken into account. Conversely, if the influence of the particle on the OP dynamics is neglected, the system is inherently out of equilibrium and we call here the tracer \textit{passive} -- this case may alternatively be regarded as an \textit{active} particle driven by temporally correlated noise.

Our method is particularly adapted to the case of a particle coupled to a strongly correlated and confined medium fluctuating on a fast (but still non-vanishing) time scale. In contrast to the approach of Refs.\ \cite{theiss_systematic_1985, theiss_remarks_1985}, which employs a quantum mechanical operator formalism, our method works directly in the space of the actual dynamical variables. Furthermore, no additional assumptions on the steady state distribution are required in our case. 

The main outcome of our method (see \cref{sec_elimination}) is a Fokker-Planck equation [\cref{eq:effective_FP}] which describes the effective Markovian tracer dynamics characterized by space-dependent drift and diffusion coefficients $\mu(R)$ and $D(R)$. The latter have been computed here up to their lowest non-trivial order in the adiabaticity parameter $\chi$ for the various cases discussed above, \ie, linear/quadratic field-particle coupling, dissipative or conserved dynamics, and passive/reactive tracer [see \cref{eq:coefficients_linear_reactive,eq:coefficients_quadratic_reactive,eq:coefficients_quadratic_reactive_bf,eq:coefficients_linear_passive,eq:coefficients_quadratic_passive}]. 
We also emphasized (when relevant) the effects of including boundary fields in the Hamiltonian of \cref{eq_Hamilt} (\ie, $h_1\neq 0$).
Our Fokker-Planck-based approach has allowed us to obtain, in a straightforward way, the effective equations of motion of a quadratically coupled tracer, including the spurious drift and higher-order corrections to the transport coefficients. We remark that an analysis based solely on the Langevin equation does in general not provide the stochastic interpretation of the effective tracer noise, except if the noise correlations turn out to be independent of position, as is the case in the bulk \cite{dean_diffusion_2011,demery_perturbative_2011,demery_diffusion_2013}. 

In \cref{sec_applications} we applied our method to media described by Landau-Ginzburg or Gompper-Schick type Hamiltonians, to which the tracer particle is coupled via linear or quadratic (non-derivative) terms of the OP. We obtained analytic expressions for the stationary distribution of the particle position and for the spatially dependent drift and diffusion coefficients (see \cref{fig:LG_quad,fig:LG_lin,fig:GS_quad,fig:GS_lin}).
Notably, the spatial dependence is a consequence of the confinement of the correlated medium and does not occur in the bulk \cite{Venturelli_2022}.
The diffusivity is typically strongly influenced by the presence of a dynamical conservation law, as is the stationary distribution of a passive tracer. The consistency of our findings with those obtained in the bulk limit $L\to\infty$ in Refs.~\cite{dean_diffusion_2011,demery_perturbative_2011,demery_diffusion_2013} has been finally confirmed in \cref{app:bulk}.

Our method may find application in the description of lipid membranes or microemulsions \cite{reister_lateral_2005, reister-gottfried_diffusing_2010, camley_contributions_2012, camley_fluctuating_2014, Stumpf_2021,Gompper_1994,Hennes_1996,Gonnella_1997,Gonnella_1998}, as well as colloidal particles in contact with a near-critical fluid medium \cite{demery_drag_2010, dean_diffusion_2011, demery_thermal_2011, demery_perturbative_2011, demery_diffusion_2013, fujitani_fluctuation_2016, Fujitani_2017,  maciolek_collective_2018, gross_dynamics_2021, Venturelli_2022, Basu_2022, Venturelli_2022_2parts}. In the latter case, a rapid OP field dynamics is obtained as a result of spatial confinement or a finite correlation length, which correspond to the typical experimental conditions \cite{Martinez_2017,magazzu_controlling_2019}.
Future extensions of the present work should address the OP field dynamics beyond the Gaussian approximation, which could in principle be obtained by analyzing the nonlinear terms $\sim \phi^n$ within a suitable weak-coupling expansion. The same level of analytical complication is entailed by the inclusion of field-particle couplings higher than quadratic in the Hamiltonian of \cref{eq_Hamilt}, since both translate into nonlinearities in the Fokker-Planck equation for the OP modes [see \cref{eq:FPE_modes}]. 
Since the statistics of the critical Casimir force can be extracted from the tracer distribution function, the present approach could provide further insights into the dynamics of the critical Casimir force \cite{dean_non-equilibrium_2009, dean_out--equilibrium_2010, gambassi_critical_2006, furukawa_nonequilibrium_2013, rohwer_transient_2017, gross_surface-induced_2018, gross_dynamics_2019} and of its fluctuations \cite{gross_CCFfluct_2021}, as well as on associated many-body effects \cite{mattos_many-body_2013, hobrecht_many-body_2015, maciolek_collective_2018, zakine_spatial_2020,squarcini_critical_2020}.
Finally, the extension of our results to higher spatial dimensions ($d=2$ or $d=3$) appears to be straightforward \cite{gross_dynamics_2021} and is very relevant for experimental applications; more refined models may in that case be devised to include the effects of the hydrodynamic transport of the tracer particle and the OP field.

\begin{acknowledgements}
We thank A. Gambassi for critical reading of the manuscript. DV would like to thank F. Andreucci and L. Rossi for useful discussions. DV acknowledges support from MIUR PRIN project “Coarse-grained description for non-equilibrium systems and transport phenomena (CO-NEST)” n. 201798CZL. 
\end{acknowledgements}


\appendix

\section{Lyapunov route to the \textit{super-adiabatic} approximation}
\label{app:lyapunov}
In the reactive case, one can obtain the $\cor{O}(\chi^0)$ part of the effective FP equation for the tracer particle, \cref{eq:super_adiabatic}, without invoking the quasi-equilibrium distribution as done in \cref{par:reactive}, but instead directly from the adiabatic elimination equations \eqref{eq:adiabatic} derived in \cref{sec_elimination}. We start from \cref{eq:Q1}, which gives to lowest order in $\chi$
\begin{equation}
    B_{nm} q\ione_m = s_n Q\izero \;\;\; \implies \;\;\; q\ione_n = Q\izero B_{nm}^{-1}s_m = Q\izero \Gamma_{nm}^{-1}\tau_m \, ,
\end{equation}
where the very last step holds in the reactive case with $\zeta=1$ -- see \cref{eq:def_B_reactive,eq_sn_def}. Next,
from \cref{eq:Q2} we infer
\beq
    B_{nj} q\itwo_{jm} + B_{mj} q\itwo_{jn} =   s_n q_m\ione +s_m q_n\ione + 2 L_n \delta_{nm} Q\izero \, ,
    \label{eq:lyap_q2}
\eeq
where in general the matrix $B_{nj}$ is non-diagonal because of the $\cor{O}(c)$ terms (see its definition in \cref{eq:def_B}). We recognize in \cref{eq:lyap_q2} a matrix Lyapunov equation in the form
\beq
    BX+XB^T=C \, ,
    \label{eq:lyap}
\eeq
with $X_{ij}=q\itwo_{ij}$,
and we need to search for a symmetric solution $X=X^T$. Such a solution is unique whenever the whole spectrum of the matrix $B$ has a definite sign \cite{bellman_1997}. Under this symmetry assumption (which will be checked below), we rewrite \cref{eq:lyap} as
\beq
    BX+(BX)^T=C.
\eeq
Since $C$ is a symmetric matrix in our case, we deduce that the solution should read
\beq
    BX = \frac12 C +\cor{A} \, ,
\eeq
where $\cor{A}$ is an anti-symmetric matrix. Choosing $\cor{A}= 0$, we obtain
\beq
    X = \frac12 B^{-1} C \, ,
    \label{eqq_lyap_sol}
\eeq
which indeed is a symmetric matrix owing to the non-trivial property $CB^T=BC$ holding in our case. We thus identify \cref{eqq_lyap_sol} as the solution we are searching for. This coincides with $q_{ij}\itwo$ given in \cref{eq:wickthm}, but we did not have to resort to Wick's theorem in order to obtain it (in particular, it holds true also for $T_R\neq T_\phi$).

Now we turn to the spectrum of the matrix $B$. From the theorems on the Sylvester equation \cite{bellman_1997}, it is sufficient to prove that the matrices $B$ and $-B$ have no common eigenvalues in order for the matrix equation \eqref{eq:lyap_q2} to admit a unique symmetric solution $X_{ij}$. For $c=0$ this is trivially true, while for $c\neq 0$ one can give an argument akin to the non-crossing rule in condensed matter physics \cite{Ashcroft_76}: indeed, the perturbation $c_{ij}$ to the matrix $B_{ij}$ is a function of the parameter $R$, so that any 
``crossing'' between eigenvalues can only be accidental and does not provide additional solutions which are valid for any choice of $R$.

\section{Reactive case with both linear and quadratic couplings}
\label{app:mixed_case}
In this Appendix we address the most general case in which both a linear and a quadratic coupling are included in the Hamiltonian in \cref{eq_Hamilt} together with boundary fields, so that $c$, $h$, $h_1\neq 0$. The calculation runs similarly to \cref{sec_quad_reac_bound} at the cost of a slight proliferation of new terms, so that the effective FP equation takes again the form of \cref{eq:effective_FP} upon defining
\beq 
    \mu(R) = \mu_0(R) + D'(R),\qquad \mu_0(R) = - \left[ 1-\chi T M_5(R)\right] \left[U'(R)+W'(R)\right], \qquad
    D(R) = T - \chi T  M_5(R) \, .
\eeq 
The function $M_5(R)$ reads
\begin{align}
    M_5(R) \equiv M_3(R) -\frac{2T F_6}{1+V} \left(G_3- \frac{G_2 G_4}{1+V}\right) +2 \left( F_1 -F_7 G_6 + \frac{G_2 G_6 F_4-F_2 G_2}{1+V} \right) \, ,
    \label{eq:M5(R)}
\end{align}
where $M_3(R)$ was given in \cref{eq:M3(R)}, we used the definitions of the functions $F_n(R)$ and $G_n(R)$ in \cref{eq:F_G}, and we introduced
\begin{align}
    &G_2(R) \equiv  \sum_i \frac{u_i t_i}{\beta_i} ,    && G_3(R) \equiv \sum_i \frac{u_i t_i}{b_i \beta_i},   \\
    &F_1(R) \equiv T \sum_{nm} \frac{A_{nm} \tau_n t_m}{b_m\beta_n\beta_m},  && F_7(R) \equiv  \sum_{nm} \frac{A_{nm} u_n t_m}{b_n+b_m} \, .
\end{align}

\section{Comparison with previous results in the bulk}
\label{app:bulk}
In this Appendix we check the consistency of our results with the effective particle dynamics derived in Refs.~\cite{dean_diffusion_2011, demery_perturbative_2011, demery_diffusion_2013} for the same model which we described in \cref{sec_model}, but in the absence of confinement, \ie, in the \textit{bulk} limit. 
Throughout the main text, we have assumed that the OP field satisfies either Neumann or Dirichlet boundary conditions (see \cref{sec_model}), while we did not address explicitly the case of periodic \bcs.
In fact, this case is arguably less interesting than the other two, since the effective drift and diffusion coefficients $\mu(R)$, $D(R)$ of the particle become $R$-independent at leading order of the adiabatic expansion (see Ref.~\cite{gross_dynamics_2021}). 
Moreover, the case of PBCs can often be addressed by starting from the results for Neumann and Dirichlet BCs, as we will detail below.

Attempting to recover the bulk limit of the coefficients $\mu(R)$, $D(R)$ by simply sending $L\to \infty$ in the final expressions corresponding to Neumann/Dirichlet \bcs would render, in general, a wrong result. 
Heuristically, this is because only \textit{half} of the modes of the OP field present in a bulk system are retained when dealing with Neumann/Dirichlet \bcs. Indeed, a periodic function $f(z)$ on the interval $[-L,L]$ admits the expansion
\begin{equation}
    f(z) = \frac{1}{\sqrt{L}}\left[ \frac{a_0}{2} +\sum_{n=1}^\infty \left( a_n \cos \frac{n\pi z}{L} + b_n \sin \frac{n\pi z}{L}  \right)  \right]  ,
    \label{eq:expansion_sin_cos_0}
\end{equation}
where the Fourier coefficients are given as usual by
\begin{equation}
    a_n = \frac{1}{\sqrt{L}}\int_{-L}^L\d z\, f(z) \cos \frac{n\pi z}{L}  \, ,\qquad
    b_n = \frac{1}{\sqrt{L}}\int_{-L}^L\d z\, f(z) \sin \frac{n\pi z}{L}  \, .\\
\label{eq:fourier_coefficients_app}
\end{equation}
Equation~\eqref{eq:expansion_sin_cos_0} essentially contains a sum of Neumann and Dirichlet eigenmodes, as we can write, using  \cref{eq_eigenspec}:
\begin{equation}
    f(z) = \frac{1}{\sqrt{2}}\left[ \sum_{n=0}^\infty a_n \sigma_n\Nbc (z) +\sum_{n=1}^\infty b_n \sigma_n\Dbc (z)   \right] \equiv \sum_{n=-\infty}^\infty c_n \sigma_n\Pbc(z)\, , 
    \label{eq:expansion_sin_cos}
\end{equation}
where $\sigma_n\Pbc(z)$ was introduced in \cref{eq_eigenf_periodic_real}, while
\beq
    c_n \equiv \begin{cases}
                          a_{-n} ,& n=0,-1,-2,\ldots \\
                          b_n ,& n=1,2,3,\ldots
        \end{cases}
\eeq

With this in mind, we now turn to the comparison with previous bulk results by starting with the \textit{linearly} coupled case. The effective particle dynamics has been obtained in the adiabatic limit in Ref.~\cite{dean_diffusion_2011} in the form of a Langevin equation.  
While the associated bulk drift term has been found to vanish, $\mu_b(R)=0$, the diffusion coefficient $D_b(R)$ takes a nontrivial form as reported in Eqs.~(22) and (23) therein. Upon expressing $D_b(R) = T - \chi T  M_b(R)$ as we did in \cref{eq:coefficients_linear_reactive}, and by calling $\chi \equiv \kappa / \kappa_\phi $ the adiabaticity parameter (\ie, the ratio of the particle/field mobilities in the notation of Ref.~\cite{dean_diffusion_2011}), the bulk result reads
\begin{equation}
        M_b(R) = (2\zeta -1) \frac{h^2}{d} \int \frac{d^d q}{(2\pi)^d} \frac{q^2 \tilde{ \mathcal{K}}_1^2(q)}{\tilde{ \Delta} (q) \tilde{ \Lambda} (q)}
        \label{eq:M_lin_dean}
\end{equation}
up to $\mathcal{O}(\chi)$.
This is valid in any dimension $d$, and the tilde stands for the Fourier transform of the operators introduced in \cref{sec_model}. Since $\zeta=0/1$ for a passive/reactive tracer respectively, the correction to the diffusion coefficient is the same in these two cases, but with the opposite sign (in particular, diffusion is enhanced in the passive case and hindered in the reactive case). Note that the same result, \cref{eq:M_lin_dean}, can be recovered by taking the \textit{adiabatic} limit $\kappa_\phi \gg \kappa$ in Eq.~(43) of Ref.~\cite{demery_perturbative_2011}.

Comparing these with our results, we note the following points:
\begin{itemize}
    \item The correction to the diffusion coefficient $M_b(R)$ in \cref{eq:M_lin_dean} reduces, for $d=1$, to $M(R)$ given in \cref{eq:M(R)} for a reactive tracer, provided that in the latter one replaces the term $\left[ \pd_R v_n(R)\right]^2$ by $|\pd_R v_n(R)|^2$, and chooses plane waves $\sigma_n(z) = \exp(\im k_n z)/\sqrt{2L} ,\, k_n = \pi n / L ,\, n \in \mathbb{Z}$ as the eigenbasis.
    In this way, the $R$-dependence evidently drops out of the integral over $q$. [Note, however, that this prescription is equivalent to choosing real periodic eigenfunctions as in \cref{eq_eigenf_periodic_real}]. The equivalence between the two expressions in \cref{eq:M(R),eq:M_lin_dean} can then be recognized by replacing the Fourier transforms of the operators $\Delta,\Lambda,\mathcal{K}_1$ introduced in \cref{sec_model} by their corresponding Fourier coefficients $\beta_n,L_n$ and $v_n$ [see \cref{sec:mode_expansion}], and the integral by a sum according to $ \int_{\mathbb{R}} d\, q \to \frac{1}{2L} \sum_{n\in \mathbb{Z}} $. Recall that the same function $M(R)$ controls the diffusion coefficient also in the passive case [see \cref{eq:coefficients_linear_passive}].
    \item The vanishing bulk drift term $\mu_b(R)=0$ is consistent with the flattening of the stationary effective potentials and of the diffusion coefficient in the bulk limit, i.e., $V(R),W(R),D(R)\to \const$ for $L\to\infty$. Indeed, the drift coefficient $\mu(R)$ is generally proportional to their derivative with respect to $R$ [see \cref{eq:super_adiabatic,eq:coefficients_linear_reactive,eq:coefficients_quadratic_reactive,eq:coefficients_linear_passive,eq:coefficients_quadratic_passive}]. We have already noted in \cref{sec_discussion} that $V(R),W(R)$ defined in \cref{sec_stationary_potentials} must become $R$-independent in the bulk by translational invariance (this has been checked explicitly in \cref{par:application_LG} for the off-critical LG model). At the critical point, $V(R)$ and $W(R)$ may reduce in the bulk limit to a structureless, IR diverging constant, which, however, does not affect their derivatives and thus the drift coefficient.
\end{itemize}

To be more concrete, let us analyze the case of the Gaussian LG Hamiltonian with model A dynamics addressed in Ref.~\cite{dean_diffusion_2011}. 
In the limit $L\to \infty$, our expression in \cref{eq:M(R)_LG_A} becomes
\begin{equation}
    M_{D/N}(R) \to \frac{h^2}{4} \left[ \xi \pm e^{-2 R/\xi} \left(\xi-2R\right)  \right] ,
    \label{eq:diff_LG_A_bulk_us}
\end{equation}
where the $\pm$ sign corresponds to Dirichlet/Neumann \bcs, respectively. 
Using \cref{eq:expansion_sin_cos} together with the definition of $M(R)$ in \cref{eq:M(R)}, we obtain the corresponding bulk result as
\beq
    M_b(R) = \frac12 [M_N(R)+M_D(R)] = \frac{h^2 \xi}{4} \, ,
\eeq
which coincides with the expression reported in Eq.~(37) in Ref.~\cite{dean_diffusion_2011} (with $d=1$ and $m\equiv 1/\xi$).

We remark that, for the LG Hamiltonian with model B dynamics, the adiabatic limit and the bulk limit are incompatible (see also Ref.~\cite{Venturelli_2022} for further details). The reason is that in model B a continuum of slow OP modes builds up at the wavenumber scale $q\sim 1/L$ when approaching the bulk limit, so that the OP field can never be considered \textit{fast} (while in model A even the slowest mode has a finite relaxation time, as long as the correlation length $\xi$ remains finite). 
This is a direct consequence of the conservation of the OP field in model B dynamics. In our formalism, this translates into the divergence of the effective adiabaticity parameter $\tilde \chi$ in \cref{eq_adiab_param} in the bulk limit, being $d_\Lambda = -2$ for model B (while $d_\Lambda =0$ for model A). 

Let us finally address the \textit{quadratically} coupled case. The correction to the diffusion coefficient has been obtained in the bulk under the \textit{weak-coupling} approximation in Ref.~\cite{demery_diffusion_2013}, see Eq.~(66) therein. Its \textit{adiabatic} limit can again be recovered by inspecting the limit for $\kappa_\phi \gg \kappa$, which renders
\begin{equation}
        M_b(R) = (2\zeta -1) \frac{h^2}{2d} \int \frac{d^d q\, d^d q}{(2\pi)^{2d}} \frac{(\qv+\pv)^2 \tilde{ \mathcal{K}}^2_2(\qv) \tilde{ \mathcal{K}}_2^2(\pv) }{\tilde{ \Delta} (\qv) \tilde{ \Delta} (\pv) \left[  \tilde{ \Lambda} (\qv)\tilde{ \Delta} (\qv) + \tilde{ \Lambda}(\pv)\tilde{ \Delta} (\pv)  \right]  }\, .
    \label{eq:M_quad_dean}
\end{equation}
This again compares very well with the correction to the diffusion coefficient presented in \cref{eq:M4(R)} for the passive case, upon replacing $A_{nm}^2$ by $|A_{nm}|^2$ and choosing plane waves for the eigenmodes (while we have shown in \cref{sec_discussion} that the correction in the reactive case reduces in the bulk limit to that of the passive case, up to a minus sign).
No explicit forms have been obtained for the diffusion coefficient of specific models in the quadratic case (either in this manuscript or in Ref.~\cite{demery_diffusion_2013}), but we have still checked their overall qualitative agreement (see \cref{sec_applications}).

\section{Effective noise in the passive-quadratic case}
\label{par:CLT}
In the passive case and in the absence of linear couplings or boundary fields, the dynamics of the field modes $\phi_n$ given in \cref{eqq_langevin_modes} reduces to
\begin{equation}
    \pd_t \phi_n =- \chi^{-1} b_n \phi_n  + \chi^{-1/2}\xi_n \, ,
\end{equation}
where the correlations of $\xi_n$ are given in \cref{eq_phi_noise_mode_correl}. At long times we thus have
\begin{equation}
    \bra \phi_n(t) \phi_m(t') \ket = \delta_{nm} \frac{L_n}{b_n} e^{-b_n|t-t'|/\chi} = \delta_{nm} \frac{T_\phi}{\beta_n} e^{-b_n|t-t'|/\chi} \, . 
\end{equation}
Similarly, the Langevin equation \eqref{eqq_langevin_modes_R} for the passive tracer can be cast in the form
\begin{equation}
    \pd_t R(t) = - \sum_{nm} A_{nm}(R) \phi_n(t)\phi_m(t)  + \eta(t) \equiv - \sum_n \varphi_n(R,t) + \eta(t) \equiv \Pi_c(R,t)+ \eta(t) \, ,
\end{equation}
where following Ref.~\cite{gross_dynamics_2021} we introduced the effective noise
\begin{equation}
    \Pi_c(R,t) \equiv - \sum_n \varphi_n(R,t) \equiv - \sum_{nm} A_{nm}(R) \phi_n(t)\phi_m(t) \, .
\end{equation}
Using Wick's theorem, it is simple to show that
\begin{align}
    \bra \varphi_n(t) \ket &= A_{nm}(R) \frac{T_\phi}{\beta_n} \, , \\
    \bra \varphi_n(t) \varphi_k(t') \ket_c &= A_{nk}A_{kn} \frac{T_\phi^2}{\beta_n\beta_k} e^{-(b_n+b_k)|t-t'|/\chi} + \delta_{nk} \frac{T_\phi^2}{\beta_n} \sum_m \frac{A_{nm}^2}{\beta_m} e^{-(b_n+b_m)|t-t'|/\chi} \, ,
    \label{eq:effective_modes_correlations}
\end{align}
whence
\begin{equation}
    \bra \Pi_c(R,t) \Pi_c(R,t') \ket_c = T_\phi^2 \sum_{nm} \frac{A_{nm}^2+A_{nm}A_{mn}}{\beta_n\beta_m} e^{-(b_n+b_m)|t-t'|/\chi} \, .
\end{equation}
By comparing the latter with the effective diffusion coefficient $D(R)$ in \cref{eq:coefficients_quadratic_passive}, we finally obtain the Green-Kubo relation in \cref{eq:Kubo}.

\section{Details of the calculation of the stationary potentials}
\label{app:details_series}
Here we give further details on the derivations presented in \cref{sec_applications}.
Let us start from the LG model, and consider the computation of $V(R)$ in the quadratic case [see \cref{eq:V(R),eq:V(R)_LG_pure}]: if $0\leq x \leq \pi$, we can use the relations \cite{gradshteyn_table_2014}
\begin{align}
    \sum_{k=1}^\infty \frac{\cos^2(kx) }{k^2+\alpha^2} &= \frac{\alpha \pi \,  \T{csch}( \alpha \pi ) \left[\cosh (\alpha \pi) + \cosh (\alpha(\pi-2x)) \right]-2}{4\alpha^2} \, , \label{eq:series_cos} \\
    \sum_{k=1}^\infty \frac{\sin^2(kx) }{k^2+\alpha^2} &= \frac{\pi \,  \T{csch}(\alpha \pi) \sinh[ \alpha (\pi - x)] \sinh(\alpha x)}{2 \alpha} \, , \label{eq:series_sin}
\end{align}
and identify  $\alpha x \to x/\xi $, $\alpha \pi \to L/\xi $.

Whenever the result is still a convergent series, we can compute derivatives as
\begin{equation}
    \sum_{k=1}^\infty \frac{\cos^2(kx) }{(k^2+\alpha^2)^2} = -\frac{\partial }{\partial (\alpha^2)} \sum_{k=1}^\infty \frac{\cos^2(kx) }{k^2+\alpha^2} \, ,
    \label{eq:relation_for_M(R)}
\end{equation}
allowing us to make use of known results. 
In this way, a closed expression for $M(R)$ in the linearly coupled case can be obtained: specializing \cref{eq:M(R)} to the LG model with linear coupling, one has
\beq
    M(R) = \frac{2 h^2}{L}  \sum_n \frac{k_n^2 \left[  \cos / \sin(k_n R)\right]^2}{(k_n^2+\tau)^2 k_n^{2a}} \, ,
\eeq
for Dirichlet/Neumann \bcs, respectively.
For model B ($a=1$), this reduces to the series in \cref{eq:relation_for_M(R)}, yielding \cref{eq:M(R)_LG_B} in the case of Neumann \bcs, and
\begin{align}
    M(R)_{\T{Dir}} =& \, \left[\frac{h\xi e^{L/\xi}}{2(e^{2L/\xi}-1 )}\right]^2  \Bigg\lbrace \frac{4}{L} \Bigg[ L^2 + 2\xi^2 +LR \cosh{\frac{2}{\xi}(L-R)}+L(L-R) \cosh{\frac{2R}{\xi}} \Bigg] \n\\
    &+ 2\xi \Bigg[\sinh{\frac{2}{\xi}(L-R)} +\sinh{\frac{2R}{\xi}}+\sinh{\frac{2L}{\xi}} -\frac{4\xi}{L}\cosh{\frac{2L}{\xi}} \Bigg] \Bigg\rbrace \label{eq:M(R)_LG_B_Dir}
\end{align}
in the case of Dirichlet \bcs (reported here mainly for formal reasons, since model B dynamics with a globally conserved OP field is incompatible with Dirichlet BCs -- see \cref{par:examples}). For model A ($a=0$), we can add and subtract $\tau$ at the numerator to write
\beq
    M(R) = \frac{2 h^2}{L} \left\lbrace \sum_n \frac{ \left[  \cos / \sin(k_n R)\right]^2}{(k_n^2+\tau)} -\tau \sum_n \frac{ \left[  \cos / \sin(k_n R)\right]^2}{(k_n^2+\tau)^2 }\right\rbrace \, .
\eeq
The first term is identical to \cref{eq:series_cos,eq:series_sin}, while the second is analogous to model B.
The two series reported above in \cref{eq:series_cos,eq:series_sin} start from $n=1$, so the zero mode $n=0$ has to be added by hand when considering Neumann BCs -- see \cref{eq_eigenf_Nbc}. Overall, this gives
\begin{align}
    M(R)_{\T{Dir}} &= h^2 \xi \left\lbrace \frac{1}{2}\, \T{csch}(L/\xi) \left[ \cosh (L
    /\xi)  + \cosh \left(\frac{L-2R}{\xi}\right) \right] -\frac{\xi}{L} \right\rbrace - \frac{1}{\xi^2} M(R)_{\T{Dir}}^{\T{model B}}  \, , \n \\
    M(R)_{\T{Neum}} &= h^2\xi \,  \T{csch}(L/\xi) \sinh(R/\xi) \sinh\left(\frac{L-R}{\xi}\right) - \frac{1}{\xi^2} M(R)_{\T{Neum}}^{\T{model B}} \, ,
\end{align}
which simplify to the expressions given in \cref{eq:M(R)_LG_A}.\\
In order to compute the part of $W(R)$ proportional to $h_1$ [see \cref{eq:W(R)_quad,eq:W(R)_lin}], we finally make use of the Werner's formulas and the relation \cite{gradshteyn_table_2014}
\begin{equation}
    \sum_{k=1}^\infty \frac{\cos(kx) }{k^2+\alpha^2} = \frac{\pi\cosh{\alpha(\pi-x)}}{2\alpha \sinh{\alpha \pi}}-\frac{1}{2\alpha^2} \, .
\end{equation}

For the critical GS model, we make use of the known relation \cite{gradshteyn_table_2014}
\begin{equation}
    f_{2n}(x) \equiv \sum_{k=1}^\infty \frac{\cos{k x}}{k^{2n}} = \frac{(-1)^{n-1} (2\pi)^{2n}  }{2(2n)!} B_{2n}\left(\frac{x}{2\pi}\right) \, ,
\end{equation}
where $B_n(x)$ is the $n$-th Bernoulli polynomial \cite{gradshteyn_table_2014}, and $x\in [0,2\pi]$. For instance,
\begin{align}
    f_4(x) & = \frac{1}{720} (8 \pi^4 - 60 \pi^2 x^2 + 60 \pi x^3 - 15 x^4) \, , \n \\
    f_6(x) & = \frac{1}{30240}(32 \pi^6 - 168 \pi^4 x^2 + 210 \pi^2 x^4 - 126 \pi x^5 + 21 x^6) \, , \n \\
    f_8(x) & = \frac{1}{1209600} (128 \pi^8 - 640 \pi^6 x^2 + 560 \pi^4 x^4 - 280 \pi^2 x^6 + 120 \pi x^7 - 15 x^8) \, .
    \label{eq:f_n_def}
\end{align}


%

\end{document}